\documentclass[12pt,aps,pre,preprint]{revtex4-1}

\usepackage{amsmath}    % need for subequations
\usepackage{amsfonts}  % note how statements can be commented out
\usepackage{graphicx}   % for figures
\usepackage{bm}  %for bold in math mode
\usepackage{calligra}
\DeclareMathAlphabet{\mathcalligra}{T1}{calligra}{m}{n}
\DeclareFontShape{T1}{calligra}{m}{n}{<->s*[2.2]callig15}{}

\newcommand{\scripty}[1]{\ensuremath{\mathcalligra{#1}}}
\def\scrmag{\scripty{r}}

\def\scrhat{\hat{\scripty{r}}}
\def\scr2{\frac{\scrhat}{\scrmag^2}}

\def\del{\overrightarrow{\nabla}}

\DeclareMathAlphabet{\mathpzc}{OT1}{pzc}{m}{it}

\begin{document}
%\begin{frontmatter}
%\title{Thermodynamic Induction Under Isothermal Conditions: Particle Transport and Induced Diffusion} 
%\title{Thermodynamic Induction Under Isothermal Conditions: Induced Particle Transport, Electromigration and Nano-Scale Manipulation by STM} 
%\title{Thermodynamic Induction Under Isothermal Conditions: Induced Particle Transport, Order by Induction, and Nano-Scale Manipulation by STM} 
%\title{Thermodynamic Induction Under Isothermal Conditions: Order-by-Induction in Particle Transport and Entropic Trapping in Nano-Scale Manipulation by STM}
%\title{Thermodynamically Induced Particle Transport: Order-by-Induction and Entropic Trapping at the Nano-Scale} 

%\title{Thermodynamic Induction up to Second Order with Applications to Phase Front Propagation, Turing Patterns, and Galaxy Rotation}
%\title{Thermodynamic Induction up to Second Order and Nonequilibrium Phase Transitions}
%\title{On the Thermodynamics of Bifurcations and Nonequilibrium Phase Transitions}
%\title{On the Thermodynamics of Nonequilibrium Phase Transitions}
%\title{Nonequilibrium Phase Transitions and the Thermodynamics of Bifurcations}
%\title{Nonequilibrium Phase Transitions as a Consequence of Second Order Thermodynamic Induction}
\title{Nonequilibrium Phase Transitions and Pattern Formation as Consequences of Second Order Thermodynamic Induction}

\author{S. N. Patitsas}
%\ead{steve.patitsas@uleth.ca}
%\thanks{}
\address{University of Lethbridge,\\4401 University Drive, Lethbridge AB, Canada, T1K3M4}

%\begin{keyword}
%nonequilibrium thermodynamics \sep Onsager relations \sep Turing patterns \sep thermodynamic induction \sep turbulence \sep convection \sep entropic trapping \sep electromigration \sep atomic manipulation
%\end{keyword}
%\affiliation{University of Lethbridge,\\4401 University Drive, Lethbridge AB, Canada, T1K3M4}


\date{\today}

\begin{abstract}

Development of thermodynamic induction up to second order gives a dynamical bifurcation for thermodynamic variables and allows for the prediction and detailed explanation of nonequilibrium phase transitions with associated spontaneous symmetry breaking.  By taking into account nonequilibrium fluctuations, long range order is analyzed for possible pattern formation.  Consolidation of results up to second order produces thermodynamic potentials that are maximized by stationary states of the system of interest.  These new potentials differ from the traditional thermodynamic potentials.
In particular a generalized entropy is formulated for the system of interest which becomes the traditional entropy when thermodynamic equilibrium is restored.  This generalized entropy is maximized by stationary states under nonequilibrium conditions where the standard entropy for the system of interest is not maximized.  These new nonequilibrium concepts are incorporated into traditional thermodynamics, such as a revised thermodynamic identity, and a revised canonical distribution.   Detailed analysis shows that the second law of thermodynamics is never violated even during any pattern formation, thus solving the entropic coupling problem.
%A theoretical treatment for a bifurcation is presented which makes the bifurcation equivalent to a nonequilibrium phase transition.  Thermodynamic induction is developed to second order and now takes into account spontaneous symmetry breaking.  Applications of this general result include a nonequilibrium generalization of the fluctuation dissipation theorem, 
%and establishment of a new thermodynamic potential that generalizes the entropy and is maximized for a system of interest well away from equilibrium.  Also, the important entropy coupling problem is solved, i.e., a detailed explanation for how the dynamical reservoir plays its role in allowing a system of interest to lower its entropy without violating the second law of thermodynamics.  
Examples discussed include pattern formation during phase front propagation under nonequilibrium conditions and the formation of Turing patterns.  The predictions of second order thermodynamic induction are consistent with both observational data in the literature as well as the modeling of this data.

\end{abstract}
%\end{frontmatter}


\maketitle %\maketitle must follow title, authors, abstract and \pacs

\section{Introduction} \label{sec:intro}

%percolation: \cite{Kirkpatrick1973}

%driven diffusive systems \cite{SCHMITTMANN1998}

%complexity \cite{Kadanoff1999}

%pattern formation review: \cite{Langer1999}

%emergent properties: \cite{Anderson1972}

%nonequilibrium phase transitions:  \cite{HINRICHSEN2006}

%roughening transition \cite{Mukamel1998}

%KPZ \cite{KPZ1986}

%examples of noneq PT \cite{Evans2008}

%self-organized criticality \cite{Wiesenfeld1987,Fogedby1989,Nagel1992}

%living systems: \cite{Munoz2018}

%earthquakes: \cite{Chakrabarti2012}

Thermodynamic induction (TI) has recently been put forward as a general approach towards the study of nonequilibrium systems and has been used to explain some important particular details regarding the manipulations of atoms and molecules by STM~\cite{Patitsas2014,Patitsas2015}.  A new type of thermoelectric cooling by TI has also been studied and proposed as a good test for the existence of TI~\cite{Patitsas2016}.  In short, TI can result in a thermodynamic variable being influenced in a surprising way when it plays the role of a gate, or control, variable.  The variable may be pushed  away from equilibrium when there is no apparent force to do so, thus giving the appearance of violating the second law of thermodynamics (SLT).  The direction of this influence is always in such a way that facilitates the approach to equilibrium of the entire system.  When the entire systems reaches equilibrium, the influence disappears.

So far, the TI theory applies to the case of a conductance coefficient (or kinetic coefficient) that depends on a thermodynamic variable in a linear fashion, i.e., the case of first order TI (TI1), ex. the electrical conductivity of a channel depending on the temperature of the channel~\cite{Patitsas2016}.  What is missing in the theory is a treatment of second order TI (TI2).  This step is important in this development of nonequilibrium thermodynamics as it will allow treatment of spontaneous symmetry breaking, pattern formation, as well as a description on nonequilibrium phase transitions (PT).  This approach also allows for a thermodynamic way view of bifurcations, a phenomenon usually approached in the realm of pure, zero temperature, mechanics.

By establishing TI up to second order, I will be able to answer a very old and important scientific question, which I term the \textit{entropic coupling problem}.  Since the establishment of the laws of thermodynamics, it has been noted that many systems in nature are highly ordered and seem to break the second law of thermodynamics.  The often given explanation is simply that even though entropy might decrease in a given region, somewhere else the entropy must increase by at least that much.  This is surely the case but a rigorous theory for this has proved elusive, until now.  In fact, to my knowledge, no theoretical work on this problem exists, beyond the qualitative explanation just given.

There must exist some other system that increases its entropy by at least as much, and this must be true at all times.  This means that the entropy production of this system must always exceed any negative rate that may occur in the given region.  Here I explain both what this other system is, as well as the details of the coupling.
%	There are two important questions.  1) What exactly is this other system? and 2) How are these two systems coupled?
%	This problem is surely important enough to deserve a name. 
%I refer to this general problem as the \textit{entropic coupling problem}.  Because of the generality of the thermodynamics involved, I believe this is one of the most important unsolved problem in all of science.  
The approach I outline here for solving the entropic coupling problem is novel.  The coupling is not energetic in the same way mechanical systems are often coupled by adding a term to a Hamiltonian that depends on the variables of both systems.   Instead, the coupling between variables occurs through the conductance.

Much of the considerations here are under circumstances I refer to as \textit{well away} from equilibrium.  What is meant by \textit{well away} is far enough from equilibrium where kinetic and transport coefficients will have significant deviations from constancy, but not so far that destructive or catastrophic events occur, i.e., the system can be repeatedly cycled well away from equilibrium and back again.  This variation of kinetic coefficients is not merely for convenience of definition, but plays a critical role in my analysis.  As it turns out, this is a modest step beyond the basic approach of near-equilibrium thermodynamics, such as used for calculation of transport properties.  Here, one is not dealing with far-from-equilibrium physics where the concepts of equilibrium statistical mechanics break down.  In particular, the local temperature, pressure, and chemical potential are still well-defined thermodynamic parameters.

%-reversible in the sense that the system is the same after XDR is brought back down to zero.  The system can be cycled repeatedly.
%well away means not at or near, but also not to be confused with discussions of catastrophic nonequilibrium events.  Basically, .

By combining TI results at both first and second order, I construct a thermodynamic potential that is maximized when the gate is well away from equilibrium and finds itself in a stationary state.
Finding such a potential has remained an open question since the laws of thermodynamics were established.  Maximizing the entropy is not helpful because this is known to happen at equilibrium.  This new potential differs from the entropy in general but does become the entropy when the system is returned to equilibrium.  Maximizing this potential will, under certain circumstances, produce a PT, which may or may not spontaneously create interesting patterns.
%Also, it has not been clear what is the nature of a state that does maximize this function.  Presumably it is quasistationary since the system is relaxing, in time, towards equilibrium.  It must then be fast variables that adjust themselves to states that maximize the potential in question, as the whole system relaxes towards equilibrium at a slower rate. 

The time is right now to use this new potential towards the establishment of governing principles for nonequilibrium thermodynamics.  An abundant amount of data has been taken from observations on a widely varying set of nonequilibrium systems over a period of many decades now.  These systems have been described in a lengthy review article~\cite{Cross1993} as well as textbooks including Refs.~\cite{Cross,Desai} as good examples.  Moreover, a great deal of modeling has been reported on these results and much understanding has been gained from this.  In this work, I make a concerted effort to link general TI results to this modeling.

I begin, in Sec.~\ref{sec:gentheory}, by developing a theory for thermodynamic induction up to second order for one gate, or control, variable.  This includes showing that TI2 produces a nonequilibrium PT with a well-defined order parameter.
In Sec.~\ref{sec:more} this theory is extended to the case of more than one gate variable.  This includes the description of a nonequilibrium front and pattern formation during chemical reactions.  Review and comparison is made to various important models in the literature.

\section{General Theory for Thermodynamic Induction up to Second Order }  \label{sec:gentheory}

Considered here is the coupled dynamics of two thermodynamic variables, referred to as the dynamical reservoir (DR) and the gate.  After some initial considerations with both variables on an equal footing, emphasis will then be placed on the gate variable.  The gate is the system capable of displaying interesting behaviour such as pattern formation and self-organization.  The DR is simply a thermodynamic variable with a large capacity so that when not in equilibrium, the relaxation is slow.  The relaxation of the DR is always considered as slowly varying compared to all other time scales.  In fact the DR may be held static in many systems, for example by continuously replacing/feeding in reactants into a reactor.
%The gate and SI are not quite the same, since the gate includes the SI as well as the traditional thermodynamic reservoirs such as a heat bath, particle reservoir, etc.  Though the variables under consideration pertain to the SI, when accounting for entropy changes, the entropy of the entire system must be considered which includes the DR, the SI, and the traditional reservoirs.  As I begin the analysis I will deal with the gate, and later on make the separation with the traditional reservoirs to focus on the SI.  
%The system of interest is incorporated into the gate.  The gate and   The details of this incorporation will be discussed below.
%  For this reason the gate variable will also be referred to as the SI, with GT and SI being used interchangeably below. 

The DR thermodynamic variable $x_{DR}$, normally considered as slowly approaching an equilibrium value of $x_{DR0}\equiv x_{DR}-a$, has a conjugate force $X_{DR}=-g_{DR}a$, where $g_{DR}^{-1}$ can be thought of as a generalized capacitance and would be large for this type of reservoir variable.
%, i.e., on the order of $N_{DR}/k_B$ where $N_{DR}$ is the number of particles in the DR.  
The dynamics for approaching equilibrium is described by
\begin{equation}
\frac{da}{dt}=\dot{a}=M_{DR}X_{DR}  \,.      \label{DRdyn}
\end{equation}
So far the analysis closely follows standard textbook material for nonequilibrium dynamics~\cite{Reif,mazur}.  Ordinarily, the Onsager coefficient $M_{DR}$ is considered as constant and describes strict proportionality between the flux and force, ex. Fourier's heat transfer law.
The key idea behind TI is that $M_{DR}$ is not constant, and may depend on thermodynamic variables other than $x_{DR}$. (A dependence on $x_{DR}$ creates nonlinear dynamics but fails to create the interesting coupling.) The coefficient $M_{DR}$ is assumed to depend on these other thermodynamic (gate) variables so that $M_{DR}$ can be broken into a sum of a constant term $L_{DR}$ and a variable component $W_{DR}$ which has a functional dependence on the gate variables.  For simplicity, I first consider the case where the dependence of $M_{DR}$ on these variables is very weak and negligible, except for one gate variable, $x_{GT}$.   The term $W_{DR}$ couples $x_{DR}$ and $x_{GT}$ and plays a role similar to the perturbation potential in the Hamiltonian description of mechanical systems.  The coupling does not occur through a Hamiltonian but instead through entropy production rates.  Both the DR and gate entropy production rates play an important role here.
%Where does this go?
The entropy production rate for the DR is
\begin{equation}
\sigma_{DR}= X_{DR}\dot{x}_{DR}= L_{DR}X_{DR}^2 + W_{DR}X_{DR}^2    \,.     \label{sigmaDR}
\end{equation}

For the gate I define the difference variable $b\equiv x_{GT}-x_{GT0}$, so that the conjugate force is $X_{GT}=-g_{GT} b$ and the change in entropy from equilibrium for the gate variable is
\begin{equation}
S_{GT}-S_{GT,eq}=\frac{1}{2}X_{GT} b= -\frac{1}{2}g_{GT} b^2   \,.      \label{SGT}
\end{equation}  
The $g_{GT}$ parameter sets the extent of fluctuations with $\langle b^2\rangle_0=k_B/g_{GT}$ (the brackets $\langle\rangle_0$ denoting equilibrium ensemble averaging).  Since the gate is often thought of a small system, fluctuations play an important role.
The entropy production rate for the gate which also plays an important role in the formulation of variational principles is given by
\begin{equation}
\sigma_{GT}= L_{GT}X_{GT}^2= L_{GT}g_{GT}^2 b^2     \,.     \label{sigmaGT}
\end{equation} 
I will show below that important potentials may be formed as linear combinations of $\sigma_{DR}$ and $\sigma_{GT}$.

\subsection{TI1}

In previous work I considered the case where $W_{DR}$ depends in a linear fashion on one or more gate, or control, variables~\cite{Patitsas2014,Patitsas2015,Patitsas2016}.  This resulted in a type of TI that is classified as first order, i.e., TI1.  If the TI1 gate variable is $b$ then $M_{DR,1}=L_{DR}+\gamma b$, with $\gamma$ as the TI1 coefficient, and the induction effect gives dynamics for the gate variable that is not merely dissipative but is instead given by: 
\begin{equation}
\dot{b}=\gamma X_{DR}^2 L_{GT}\tau^*   -L_{GT}g_{GT}b~\,, \,\,\,\,\,\,\,\,\,\,\,\,\,\,\,\,\,\,\,\,\,\,\,\,\,\,\, \text{(1st order, uncorrected)} \label{aqdot1st}
\end{equation}
where $\tau^*$ is the characteristic time for the fastest fluctuations in the gate variable.  For TI1 the induction term in Eq.~(\ref{aqdot1st}) is constant and this constant term pushes $b$ away from its equilibrium value of zero.

%Think about expressing the kappa1 version here, helpful for the galaxy rotation (but divide by b here?)

Before proceeding to TI2, I point out that essentially the same result, Eq.~(\ref{aqdot1st}), can be found by considering the two variables, $a$ and $b$ as random walkers.  As pointed by Wigner in his analysis of various proofs for the (linear) Onsager relations, a random walker where the result of each step is weighted by $\exp(\Delta S/k_B)$ will give a mean value that relaxes properly towards equilibrium~\cite{Wigner1954}.  For the two variables considered here, $\Delta S=-g_{DR}a^2-g_{GT}b^2$.  I introduce an interesting coupling by making the step length for variable $a$ depend on $b$.  
In Appendix A, the two random walker problem is analyzed and for linear dependence the analytic results for the mean values for $a$ and $b$ are given by
\begin{equation}
\dot{a}=-\frac{q_1}{\tau_0}\left(l_0^2+\delta b\right)a  \,, \,\,\,\,\,\,\,\,\,\,\,\,\,\,\, \text{(random walk, 1st order)}       \label{DRdynRand1}
\end{equation}
and
\begin{equation}
\dot{b}=\frac{q_1 \delta}{2\tau_0 } l_2^2\left(q_1 a^2-1\right) -\frac{q_2}{\tau_0}l_2^2 b  \,. \,\,\,\,\,\,\,\,\,\,\,\,\,\,\, \text{(random walk, 1st order)}      \label{GTdynRand1}
\end{equation}
To compare this result with Eq.~(\ref{DRdyn}) and Eq.~(\ref{aqdot1st}) one notes that $q_1=g_{DR}/k_B$, $q_2=g_{GT}/k_B$, $X_{DR}=-g_{DR}a$, $X_{GT}=-g_{GT}b$, and relaxation times, $\tau_{DR}=1/L_{DR}g_{DR}=\tau_0/q_1 l_0^2$, $\tau_{GT}=1/L_{GT}g_{GT}=\tau_0/q_2 l_2^2$.  This means $L_{DR}=l_0^2/k_B\tau_0$ and $L_{GT}=l_2^2/k_B\tau_0$.  The Onsager coefficients $L_{DR}$ and $L_{GT}$ are simply the random walk diffusion coefficients divided by $k_B$.  Equation~(\ref{DRdynRand1}) is the same as Eq.~(\ref{DRdyn}) as long as $\delta=\gamma k_B\tau_0$.
%$\gamma=q_1\delta/g_{GT}\tau_0=\delta/k_B\tau_0$
With this substitution Eq.~(\ref{GTdynRand1}) becomes 
%\begin{equation}
%\dot{b}=\frac{g_{DR} \gamma}{2 } L_{GT}k_B\tau_0\left((g_{DR}/k_B) X_{DR}^2/g_{DR}^2-1\right) -L_{GT}g_{GT}b  \,. \,\,\,\,\,\,\,\,\,\,\,\,\,\,\, \text{(random walk, 1st order)}      \label{GTdynRand1b}
%\end{equation}
\begin{equation}
\dot{b}=\frac{ \gamma}{2 } L_{GT}\tau_0\left(X_{DR}^2-k_B g_{DR}\right) -L_{GT}g_{GT}b  \,. \,\,\,\,\,\,\,\,\,\,\,\,\,\,\, \text{(random walk, 1st order)}      \label{GTdynRand1b}
\end{equation}
Thus, it is natural to identify the random walk time step as twice $\tau^*$.  This gives 
\begin{equation}
\dot{b}=\gamma  L_{GT}\tau^*\left( X_{DR}^2-\langle X_{DR}^2\rangle_0 \right)   -L_{GT}g_{GT}b~\,, \,\,\,\,\,\,\,\,\,\,\,\,\,\,\,\,\,\,\,\,\,\,\,\,\,\,\, \text{(1st order, corrected)} \label{aqdot1stcorr}
\end{equation} 
which closely resembles Eq.~(\ref{aqdot1st}) except for the extra term $\langle X_{DR}^2\rangle_0=k_B g_{DR}$ which represents the variance of the DR force under equilibrium conditions.  The extra term guarantees that the mean value of $\dot{b}$ is zero in equilibrium, as it should be.  Evidently the random walk analysis is more accurate than the derivation of Eq.~(\ref{aqdot1st}) presented in Ref.~\cite{Patitsas2014}.  When the DR is large then fluctuations play less of a role and the correction term is small.  The assumption made in Ref.~\cite{Patitsas2014} is that the DR is very large and slow, so it makes sense that terms like $\langle X_{DR}^2\rangle_0$ are missed in this analysis.  In contrast, Eqs.~(\ref{DRdynRand1}) and (\ref{GTdynRand1}) hold regardless of how large and slow the DR is compared to the gate.

%\includegraphics[width=6in]{Figzigm6.eps}
%\centerline{\includegraphics[width=10.0cm]{FigTI1.png}}
\begin{figure}[ht]%
\includegraphics[width=0.6\columnwidth]{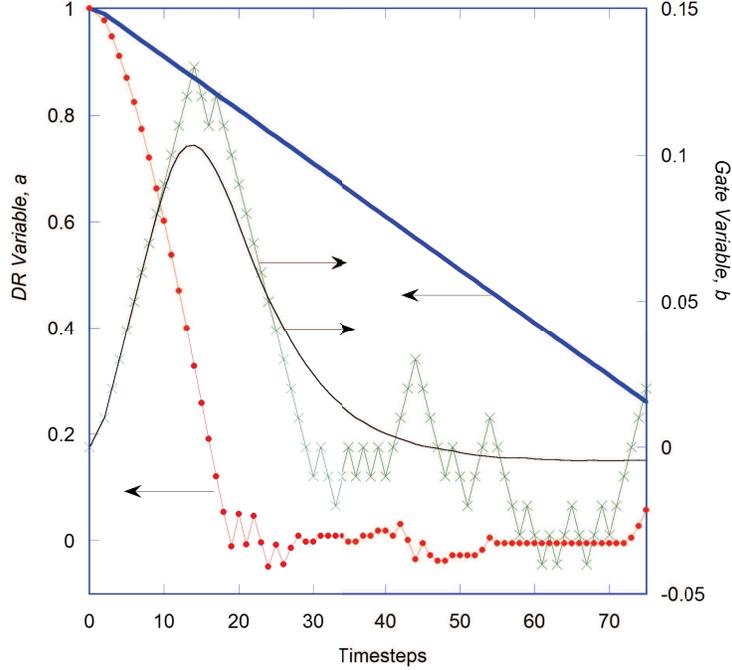}%
\caption{Random walk simulations for 1st order TI.  A single random walk is shown with 1st order TI parameter $\delta=0.04$, showing the DR (GT) variable $a$, solid red circles ($b$, green crosses) initially at 1.0  (0.0).  An average for $b$ over 10000 walks is displayed as the thin solid black curve.  Also shown is the relaxation of $a$, thick solid blue curve, when $\delta=0$, $a$ and $b$ uncoupled.  Both parameters $q_1$ and $q_2$ were set to 1000.}%
\label{FigTI1}%
\end{figure}

Figure~\ref{FigTI1} shows the result of a single, representative, random walk (solid red circles for $a$, green crosses for $b$) which clearly shows variable $b$ being pushed away from zero.  In the numerical simulation, which uses very simple code, it is the square of the DR step length that has the form $l_0^2+\delta b$.  In the simulation $l_0=0.01$, and $\delta=0.04$.  The thin, black, solid curve shows $b$ after being averaged over 10000 walks.  After about 20 timesteps $a$ relaxes to zero so that the induction on $b$ is greatly reduced,  Afterwards, $b$ relaxes towards equilibrium.
%check the offset term? -artefact from delta being nonzero, never mind
The thick, solid, blue curve shows the averaged response of $a$ when $\delta=0$, (and hence $b=0$) i.e., with the coupling between $a$ and $b$ removed (no induction).
The approach to equilibrium for $a$ is substantially faster with $\delta=0.04$ than it is for the case where $\delta$ is set to zero. 

Though the essence of the results displayed in Fig.~\ref{FigTI1} was already established in previous work it is reassuring to see a different approach based on random walk simulations confirm the expected TI1 predictions.  This confirmation will also be displayed for TI2.  Before moving to TI2, I briefly discuss a potential that is maximized under TI1.

\subsubsection{TI1 Principle of Maximum Entropy Production}

Since a correction for DR fluctuations has been added to the dynamics, Eq.~(\ref{aqdot1stcorr}), for TI1, as compared to what was derived in Ref.~\cite{Patitsas2014}, a slightly updated form of the Principle of Maximum Entropy Production (PMEP) is required here.
 Towards this end I define the following thermodynamic potential function, a type of entropy production rate:
\begin{equation}
\Phi_1\equiv \sigma_{DR}-\lambda_1\sigma_{GT}=L_{DR}X_{DR}^2+X_{DR}^2\gamma b -\lambda_1 \left[-g_{GT}\gamma  L_{GT}\tau^*\left( X_{DR}^2-\langle X_{DR}^2\rangle_0 \right)b +L_{GT}g_{GT}^2 b^2 \right]   ~\, , \label{Phi1}
\end{equation}  
where the constant $\lambda_1$ is a Lagrange multiplier.
One maximizes the DR rate of entropy production, $\sigma_{DR}$, with respect to $b$, subject to the stationary state constraint.  Equivalently, one also maximizes $\sigma_{T}$, subject to the same stationary state constraint.   Setting the first derivative of $\Phi_1$ to zero gives a way to specify $\lambda_1$ as
\begin{equation}
\lambda_1= \frac{\tau_{GT}}{\tau^*}\frac{X_{DR}^2}{X_{DR}^2-\langle X_{DR}^2\rangle_0}   ~\, , \,\,\,\,\,  \label{lambda1}
\end{equation}
which is positive definite.  Explicitly, $\partial\Phi_1/\partial b=2 \lambda_1 g_{GT} \dot{b}$.  At the stationary state $\partial^2\Phi_1/\partial b^2=-\frac{2g_{GT}}{\tau^*}\frac{X_{DR}^2}{X_{DR}^2-\langle X_{DR}^2\rangle_0}<0$, verifying that $\Phi_1$ is maximized.
%$\partial^2\Phi_1/\partial b^2=-\frac{2L_{GT}g_{GT}^2\tau_{GT}}{\tau^*}\frac{X_{DR}^2}{X_{DR}^2-\langle X_{DR}^2\rangle_0}$
%\begin{equation}
%\frac{\partial\Phi_1}{\partial b}=2 \lambda_1 g_{GT} \frac{db}{dt}    ~\, , \label{dPhibdot}
%\end{equation} 
%an expression that will be useful below.
% In essence, the intricacies of the DR are mainly filtered out of the problem and are 

\subsection{TI2} \label{sec:SOTI}

In many systems, symmetry considerations preclude the linear form of $M_{DR}$ discussed above, and the quadratic term becomes the leading term with $W_{DR}=\alpha b^2$:
\begin{equation}
M_{DR}= L_{DR}+  \alpha b^2     \,,      \label{MDR}
\end{equation}
where again $L_{DR}$ is constant and $\alpha$ is the coefficient for TI2.  In these systems, one might expect spontaneous symmetry breaking to occur.  In TI1, the gate variable is pushed away from equilibrium in a direction determined by the sign of $\gamma$.  In TI2, the gate can potentially veer away from equilibrium in either direction, so a bifurcation may occur.  This would not occur when the DR is at or very near equilibrium so one might expect a threshold needs to be exceeded before any bifurcation occurs. 

\begin{figure}[ht]%
\includegraphics[width=0.6\columnwidth]{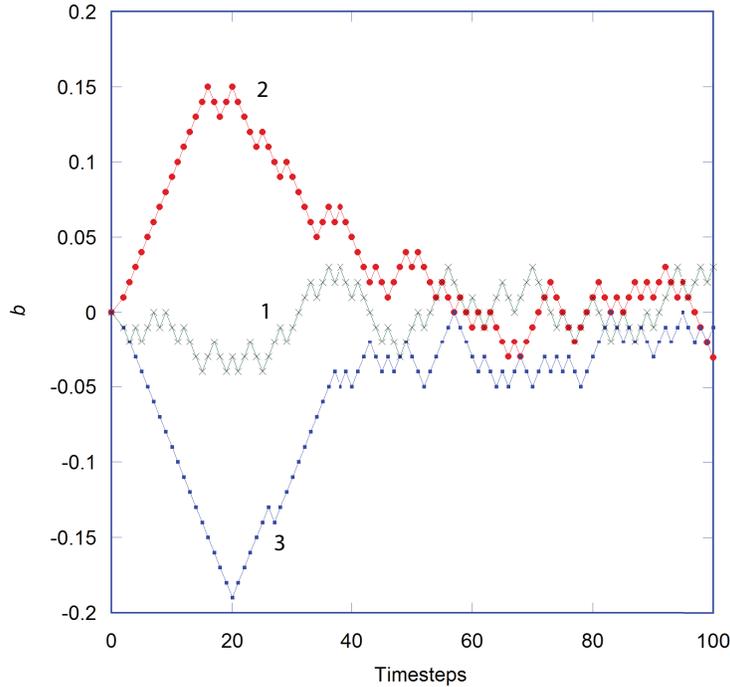}%
\caption{Random walk simulations for TI2.  Three individual random walks for the gate variable $b$ are displayed, one with $X_{DR,0}=0$ (black crosses, 1), and two with $X_{DR,0}>X_c$  (red circles, 2, and blue squares, 3).  For the random walk with $X_{DR,0}=0$, the DR $a$ variable (not shown) initial value was 0.0.  The two walks with $X_{DR,0}>X_c$ are representative of bifurcations going up and down. For these two random walks, the $a$ variable initial value was 1.0.  Both parameters $q_1$ and $q_2$ were set to 1000 and $\beta=0.1$.  Note the pitchfork shape would be evident over the range from -20 to +20 timesteps.}
\label{FigTI2}%
\end{figure}

In Fig.~\ref{FigTI2}, I present random walk results for the case where the DR step length, $l$, for variable, $a$, depends quadratically on $b$ as $l^2=l_0^2+\beta b^2$. 
Three individual random walks for $b$ are shown depicting no bifurcation (black crosses, trace 1) when the DR is at equilibrium, and two representative walks (traces 2, 3) showing bifurcation when the DR is pushed well away from equilibrium.  As in the case of TI1, the induction has significant impact on the gate variable.  Without induction the simulations always resemble trace 1.  If induction is suddenly introduced at $t=0$ then after executing a random walk for some time before $t=0$ which resembles trace 1, at $t=0$ the random walk resembles one of traces 2 and 3, and after compiling many runs, one obtains the well-known pitchfork bifurcation.  In the second order case the variable $b$ averages to zero over many random walks but as one would expect, $b^2$ does show the bifurcation.

 %When the initial position $a_0$ of variable $a$ is set to 0, both variables execute simple random walks about their equilibrium values of 0.  However, when  $a_0$ is set to 1.0, the trajectory of $b$ shows a bifurcation as shown by two representative walks.  When many random walks are executed, $b$ averages to close to zero, but $b^2$ shows the bifurcation and takes values well above the equilibrium fluctuations.

The analytic solution for the random walks is given in Appendix A as
\begin{equation}
\dot{a}=-\frac{q_1}{\tau_0}\left(l_0^2+\beta b^2\right)a  \,, \,\,\,\,\,\,\,\,\,\,\,\,\,\,\, \text{(random walk, 2nd order)}       \label{DRdynRand2}
\end{equation}
and
\begin{equation}
\dot{b}=\frac{q_1 \beta}{\tau_0 } l_2^2\left(q_1 a^2-1\right)b -\frac{q_2}{\tau_0}l_2^2 b \,. \,\,\,\,\,\,\,\,\,\,\,\,\,\,\, \text{(random walk, 2nd order)}      \label{GTdynRand2}
\end{equation}
Equation~(\ref{GTdynRand2}), with the first term on the right-hand side having a positive coefficient of $b$, shows a simple (pitchfork) bifurcation, confirming the numerical random walk results.
Making a similar mapping to the thermodynamic problem as done above with the 1st order, one obtains:
\begin{equation}
\dot{a}=  (L_{DR}+  \alpha b^2 )X_{DR}  \,, \,\,\,\,\,\,\,\,\,\,\,\,\,\,\, \text{(2nd order TI)}       \label{DRdyn2}
\end{equation}
and
\begin{equation}
\dot{b}=2\alpha L_{GT}\tau^* \left( X_{DR}^2-\langle X_{DR}^2\rangle_0 \right) b  -L_{GT}g_{GT}b~  \,. \,\,\,\,\,\,\,\,\,\,\,\,\,\,\, \text{(2nd order TI, unconstrained)}      \label{GTdyn2}
\end{equation}
Equation~(\ref{GTdyn2}) constitutes an important result of this paper as it shows how unstable dynamics may occur in a purely dissipative system.  The same result, though missing the $\langle X_{DR}^2\rangle_0$ term, is derived in a different way in Appendix B, using the same approach used in Ref.~\cite{Patitsas2014} to establish TI1.

When the force $X_{DR}$ is small, the gate variables continue to relax to equilibrium with the only effect of TI being that the relaxation time is lengthened.  
%(I don't know where to put this but the PSD for b fluctuations increases (while the BW decreases). -total noise goes as 1/g)
However when the DR is pushed harder, a critical value for $X_{DR}$ can be reached which results in a bifurcation and unstable growth of the gate variable:  
\begin{equation}
X_{c}= \sqrt{\frac{g_{GT}}{2\alpha \tau^*}+\langle X_{DR}^2\rangle_0}        ~\, . \label{Xcrit}
\end{equation}
For $X_{DR}$ fixed and above critical, Eq.~(\ref{GTdyn2}) predicts unabated growth in time of the form $b(t)=b(0)\exp(t/\tau_{})$ with
\begin{equation}
\tau^{-1} = 2\alpha L_{GT}\tau^* (X_{DR}^2-X_{c}^2)  ~\, .         \label{tauq}
\end{equation}
%Here's a good place to discuss 1st order.
The second order result expressed in Eq.~(\ref{GTdyn2}) differs significantly from TI1 which adds a constant positive term to  Eq.~(\ref{GTdyn2})~\cite{Patitsas2014}.
%Treating TI in this case has already been discussed in Ref.~\cite{Patitsas2014} with the result:
%\begin{equation}
%\dot{a}_{0}=2\alpha_0 X_{DR}^2 L_{00}\tau_0^* x_{0,eq}  -L_{00}g_{0}a_0~\, , \label{a0dot}
%\end{equation}
%where the first term on the right hand side represents first order TI.  
%The biggest difference compared to second order TI is that the first order TI term is constant in the dynamics~\cite{Patitsas2014}.  When $a_0\approx 0$ the TI pushes $a_0$ more effectively away from zero until $a_0$ settles into a stationary state where
%\begin{equation}
%a_{0,ss}=\frac{2\alpha_0 X_{DR}^2 \tau_0^* x_{0,eq}}{g_{0}}~\, . \label{a0stat}
%\end{equation}
There is no sudden transition (bifurcation) in first order TI and also no exponential growth;  the gate variable always relaxes to a nonequilibrium stationary state.
%The sign of $a_{0,ss}$ is such that both $M_{DR}$ and $\sigma_{DR}$ are always increased relative to when $a_0=0$ and $a_q=0$.  Given the form of Eq.~(\ref{sigmaDR}) one can readily see that with $a_0$ stationary, $M_{DR}\geq L_{DR}$.  Though nothing special happens to $a_0$ at  the critical point, it is worth pointing out that $a_0$ would be quite significant, and that $x_0^2/x_{0,eq}^2$ may differ from unity by an amount of unity order.

% randwkDR.csv    Set up to go in same folder as the matlab program RandWalkDR_GateAveraging.m  ; gate is 3rd column

The form of Eq.~(\ref{GTdyn2}) shows that TI can produce what is effectively a negative relaxation time.  In some systems this could happen in the form of a negative (effective) diffusion coefficient as in the physical examples discussed below in Sec.~\ref{sec:front} and Sec.~\ref{sec:turing}.  

%On its own, a system with a negative relaxation time or diffusion coefficient would seem to break the second law of thermodynamics.
%add a note here about well away, bigger than XDRrms, less than catastrophic.  
%If XC near XDRrms then this phase transition would not be easily discerned from an equilibrium phase transition.
Introducing terms leading to negative relaxation times is the basic approach in formulating the Swift-Hohenberg (SH) model~\cite{Cross1993,Cross}.  This mathematical model is quite sophisticated with extra terms added to constrain unstable solutions.  Following this approach, I add a cubic term to  Eq.~(\ref{GTdyn2}) to produce:
\begin{equation}
\dot{b}=2\alpha L_{GT}\tau^*\left(X_{DR}^2-\langle X_{DR}^2\rangle_0\right) b  -L_{GT}g_{GT}b -c^2 b^3~\,. \,\,\,\,\,\,\,\,\,\,\, \text{(2nd order, constrained)}  \label{bdot}
\end{equation}
The cubic term constrains what would otherwise be unabated exponential growth.
%For purely dissipative systems. the positive constant $g_4$, which enters into the entropy expansion, constrains what would otherwise be unabated growth.  
The physics behind the added term would have to be justified on a case-by-case basis and there may be more than one effect contributing to this cut-off. 

For example, in the realm of purely dissipative systems, treating the small $b$ expansion for $\Delta S$ to higher order will produce terms that will limit this growth.  Replacing $\Delta S=-g_{GT}b^2/2$ with
\begin{equation}
\Delta S=-\frac{1}{2}g_{2}b^2-\frac{1}{2}g_{4}b^4~\, ,  \label{g2g4}
\end{equation}
is equivalent to replacing $g_{GT}$ with $g_{2}+g_4 b^2$ with the result of producing Eq.~(\ref{bdot}) as long as $c^2=g_4 L_{GT}$.  This analysis is only preliminary and a more thorough treatment should be developed.  Nevertheless this provides a simple mechanism for limiting the unstable growth.  This connection between the entropy and the cut-off term will be invoked below in Sec.~\ref{sec:canon}.

The new term in Eq.~(\ref{bdot}), cubic in $b$, modifies the dynamics in a way that checks the exponential growth and allows for stationary states ($\dot{b}=0$) above critical:
\begin{equation}
b_{ac}=\sqrt{\frac{2\alpha L_{GT} \tau^*( X_{DR}^2-X_{c}^2) }{c^2 }}~\, . \,\,\,\,\,\,\,\, \text{(above critical)} \label{bss}
\end{equation}
Strictly speaking, this state is quasistationary and is only meaningful when the relaxation time of the gate is much shorter than that of the DR.  In this case the $X_{DR}$ is slowly changing and $b_{ac}$ adjusts according to Eq.~(\ref{bss}).  
The stationary state value of $b$ can be interpreted as an order parameter, which spontaneously springs up from zero at the critical point.  The critical exponent of this nonequilibrium PT is defined by the dependence on $X_{DR}$ just above the critical point, and here takes the value 0.5. 

Equation~(\ref{bdot}) does not account for fluctuations in $b$ and these will be dealt with below.  The results, Eqs.~(\ref{bdot}) and (\ref{bss}) are still useful in a certain limit which I refer to as the infinite $Q$ limit.  In this limit, induction does nothing when $X_{DR}<X_c$, and has a perfectly sharp transition at $X_c$.  This will change in the finite $Q$ case.

\subsection{TI2 PMEP}

In this section a principle of maximum entropy production (PMEP) will be established and will complement the same principle which holds for TI1.  For TI2 the PMEP will closely resemble the Ginzberg-Landau-Wilson free energy functional~\cite{Desai} which has been successful in modeling equilibrium PTs.  Establishing a similar thermodynamic potential should aid in understanding nonequilibrium PTs.

%As in the case of first order TI, a principle of maximum entropy production may be obtained by using the method of Lagrange multipliers, with the constraint being that the gate state is stationary.  When the gate is stationary, $\sigma_{GT}=0$ so I define the following thermodynamic potential function:
%\begin{equation}
%\Phi\equiv \sigma_{DR}-\lambda\sigma_{GT}   =L_{DR}X_{DR}^2+X_{DR}^2\alpha b^2  -\lambda \left(\alpha X_{DR}^2 L_{GT}\tau^* b^2  -L_{GT}g_{GT}b^2 -c b^4 \right)   ~\, , \label{Phi}
%\end{equation}  
%where the constant $\lambda$ is a Lagrange multiplier. 
%That $\sigma_{T}$ is larger than $L_{DR}X_{DR}^2$, i.e., the value with no TI, suggests that there may be a maximal entropy production principle for stationary states. 

\subsubsection{TI2 PMEP - Infinite $Q$}

In the case of TI2, the starting point for a PMEP potential is
\begin{equation}
\Phi_{2,\infty Q}(b)\equiv \sigma_{DR}-\lambda\sigma_{GT}=L_{DR}X_{DR}^2+X_{DR}^2\alpha b^2 -\lambda \left(-2\alpha \left(X_{DR}^2 -X_c^2\right) L_{GT}g_{GT}\tau^* b^2 +c^2 g_{GT} b^4 \right)   ~\, , \label{Phi2}
\end{equation}  
where $\lambda$ above critical is found to be    
\begin{equation}
\lambda_{2,\infty Q}= \frac{\tau_{GT}}{2\tau^*}\frac{X_{DR}^2}{X_{DR}^2-X_{c}^2}   ~\, . \,\,\,\,\, \text{(above critical)} \label{lambda}
\end{equation}
Above critical, at the stationary state, $\frac{d^2\Phi_{2,\infty Q}}{db^2} =-8\alpha X_{DR}^2$ 
%which guarantees that $\Phi_2$ is maximized at the stationary state.
%\begin{equation}
%\frac{d^2\Phi}{db^2} =-8\alpha X_{DR}^2   ~\, , \,\,\,\,\,\,\,\, \text{(above critical)}  \label{2Phia}
%\end{equation} 
which is always negative and shows that stationary states maximize $\Phi_{2,\infty Q}$.  As will be shown below, Eq.~(\ref{Phi2}) is only valid above critical in the limit of infinite quality factor. 

Above critical, $\Phi_{2,\infty Q}(b)$ resembles an upside down sombrero with two maxima at nonzero $b=\pm b_{ac}$ which in turn closely resembles the free energy functional found in Landau theory~\cite{Bergerson}.  Far enough above critical, transitions between $+b_{ac}$ and $-b_{ac}$ become rare and the system essentially freezes into one branch, i.e., a proper bifurcation in the sense of nonlinear dynamics theory.
As one varies $X_{DR}$ and approaches the critical point from above, the two maxima soften and coalesce into one maximum at $b=0$. 

%Above critical, the first derivative works out as
%\begin{equation}
%\frac{\partial\Phi_2}{\partial b}=4 \lambda g_{GT} \frac{db}{dt}    ~\, , \label{dPhi2bdot}
%\end{equation} 
%an expression that will be useful below.

\subsubsection{TI2 PMEP - Finite $Q$}

The form of $\lambda_{2,\infty Q}$ is similar to that of $\lambda_1$, with $\langle X_{DR}^2\rangle_0$ replaced by the larger $X_c^2$.  This means that the possibility of $\langle X_{DR}^2\rangle_0$ less than $X_c^2$ should be considered.
%One can also say the $\sigma_{DR}$ is maximized subject to the stationary state constraint, and one can also say that $\sigma_{T}$ is maximized subject to that same constraint.
Below critical, $\lambda$ is indeterminate and could be set to zero.  From the mathematics alone it seems reasonable to consider the function that is the magnitude of $X_1^2/(X_1^2-X_c^2+i\epsilon)$ as a function that could cover cases both above and below critical.  This can be justified by analysis that takes fluctuations in $b$ into better account.  From this analysis comes a lineshape function to replace $\lambda$ with $\lambda_2(X_{DR})=(\tau_{GT}/2\tau^*)\mu$, with $\mu$ given by:
\begin{equation}
\mu=\frac{X_{DR}^2}{\sqrt{(X_{DR}^2-X_{c}^2)^2+X_{DR}^2X_c^2/Q^2}}   ~\, , \,\,\,\,\, \text{(lineshape)} \label{mu}
\end{equation}
where the quality factor is specified by $Q^{-2}=4c^2 \tau_{GT}\langle b^2\rangle_0$.  The lineshape function $\mu$ is identical to that which comes about from the frequency response of a damped harmonic oscillator. The familiarity of this lineshape function helps to make the thermodynamics of bifurcation quite intuitive.
 Loosely speaking, the thermodynamic force $X_{DR}$ plays a role similar to frequency or energy of excitation, and $X_c$ is like the resonant frequency.  Fluctuations broaden the excitation peak and produce the interesting consequence of the system bifurcating (in a statistical manner) to some extent, below the critical point.  Unlike the case of pure classical mechanics at zero temperature where bifurcation occurs with perfect sharpness, here the bifurcation is broadened over a range $\approx X_{c}/Q$.  This constitutes an important result in the thermodynamics of bifurcations.

The lineshape function reaches below critical, suggesting that Eq.~(\ref{Phi2}) might be applicable below critical if $\lambda$ is replaced by $\lambda_2$.  If so, then dynamics below critical would be specified by making $\dot{b}$ proportional to $\partial\Phi_2/\partial b$.
However, further analysis requires care in treating fluctuations and in distinguishing $b^2$ from $\langle b^2\rangle$.  In Eq.~(\ref{Phi2}) it is the dispersion of $b$, $u\equiv\Delta b^2=\langle b^2\rangle-\langle b^2\rangle_0$, that is the argument and not the square of the mean, $\langle b^2\rangle$.  In equilibrium it is $u$ that is zero, not $\langle b^2\rangle$.   

Since focusing on fluctuations in the gate is the more important issue for TI2, I'll assume from now on that fluctuations in the DR are negligible.  This is justified by the DR being physically much larger than the gate.  
% Thus $2\alpha\tau^2=g_{GT}/X_c^2$ 
%show uss quadratic form using Q and Delta SGT
%I think I want udot here but think it over.  -will refer to it for momentum eqn
Multiplying Eq.~(\ref{bdot}) through by $b$ gives
\begin{equation}
\frac{db^2}{dt}=2b\dot{b}=4\alpha L_{GT}\tau^*\left(X_{DR}^2- X_{c}^2\right) b^2  -2c^2 b^4~\,. \,\,\,\,\,\,\,\,\,\,\,\,\,\,\,\,\,\,\,\,\,\,\,\, \text{(uncorrected)}   \label{b2}
\end{equation}
Taking into account fluctuations in $b$ means that a term $\langle b^2\rangle_0$ must be added into the dynamics for $b^2$.
This ensures that $b^2$ takes the limit $\langle b^2\rangle_0)$ well below critical.  Equation~(\ref{b2}) gets modified to become
\begin{equation}
\frac{db^2}{dt}=4\alpha L_{GT}\tau^*\left(X_{DR}^2- X_{c}^2\right) b^2   +4\alpha L_{GT}\tau^* X_{c}^2 \langle b^2\rangle_0  -2c^2 b^4~\,.   \label{b2mod}
\end{equation}

The dynamics for the difference $u=b^2-\langle b^2\rangle_0$ requires a small adaptation for the quartic cutoff term:
\begin{equation}
\frac{du}{dt}=4\alpha L_{GT}\tau^*\left(X_{DR}^2-X_c^2\right)u+4\alpha L_{GT}\tau^* X_{DR}^2\langle b^2\rangle_0  -2c^2 u^2 ~\,.   \label{b2mod2}
\end{equation}
In terms of $z\equiv X_{DR}/X_c$ the stationary state for the dispersion of $b$ is
\begin{equation}
\left(\Delta b^2\right)_{ss}=2Q^2\langle b^2\rangle_0\left[z^2-1+\sqrt{(z^2-1)^2+z^2 /Q^2}\right]    \,.   \label{uquad}
\end{equation}
In terms of the function $h_+$ defined by $h_+(x)\equiv x-1+\sqrt{(x-1)^2+x/Q^2}$, $\left(\Delta b^2\right)_{ss}=2Q^2\langle b^2\rangle_0h_+(z^2)$.  For positive argument $h_+$ is always positive and resembles a hockey-stick with a sharp upwards bend at $x=1$.  The function $h_+(z^2)$ differs very little from the infinite $Q$ limit except within the interval centered at $z^2=1$ and with a width of several $Q^{-1}$.  For small $x$, $h_+(x)\approx x/2Q^2$, while for $x\gg 1$, $h_+(x)\approx 2x$.
Far above critical,  $\left(\Delta b^2\right)_{ss}=b_{ac}^2$, showing that this approach connects seamlessly with Eqs. (\ref{Phi2}) and (\ref{lambda}) when well above critical.  This also is consistent with classical chaos theory (where $Q$ is infinite) and one expects the classical particle to strictly go to either $+b_{ac}$ or $-b_{ac}$ after bifurcation.  Near critical the level of fluctuation is substantially higher than it is in equilibrium.  This enhancement motivates defining the term, \textit{nonequilibrium fluctuations}, as an interpretation of Eq.~(\ref{uquad}). 

%Add note on h'

To describe conditions both above and below critical for TI2 one must use a potential $\Phi_2$ that is a function of $u$ since the mean value of $b$ is technically zero.  In this case the Lagrange multiplier is $\lambda_2=\tau_{GT}\mu/2\tau^*$, and the correct form for this potential is
%\begin{equation}
%\Psi_2(u)=\frac{1}{2}L_{DR}X_{c}^2\tau^* z^2+\frac{g_{GT}}{4}\left[ \mu z^2\langle b^2\rangle_0 +z^2 u +\mu(z^2-1)u-\frac{\mu}{4Q^2\langle b^2\rangle_0}u^2\right] ~\, . \label{Psi2mod}
%\end{equation}
%\begin{equation}
%\Phi_2(u)=L_{DR}X_{c}^2 z^2+\frac{g_{GT}}{2\tau^*}\left[ \mu z^2\langle b^2\rangle_0 +z^2 u +\mu(z^2-1)u-\frac{\mu}{4Q^2\langle b^2\rangle_0}u^2\right] ~\, . \label{Psi2mod}
%\end{equation}
\begin{equation}
\Phi_2(u)=L_{DR}X_{DR}^2 +\frac{g_{GT}\mu}{2\tau^*}\left[z^2\langle b^2\rangle_0 +u h_+(z^2)-\frac{u^2}{4Q^2\langle b^2\rangle_0}\right] ~\, . \label{Psi2mod}
\end{equation}
For all $u\geq 0$, $\Phi_2$ is maximized, with respect to $u$, at the stationary state Eq.~(\ref{uquad}).  As in the first order case, one maximizes $\sigma_{DR}$ (or equivalently $\sigma_T$) subject to the stationary state constraint.

The second order PMEP nicely complements the first order PMEP first discussed in Ref.~\cite{Patitsas2014}.  The second order result is quite a bit more involved;  Attempts to use the same variable for the arguments of both $\Phi_1$ and $\Phi_2$ are not successful.  In the end, it is the statistical moments of the statistical variable $b$ that are used as the arguments, i.e., $\langle b\rangle$ for $\Phi$, and $\langle u\rangle=\langle b^2\rangle-\langle b^2\rangle_0$ for $\Phi_2$.  The two moments are generated from a probability distribution.  It is helpful to think then of a Gaussian distribution for $b$.  The distribution is centered at $\langle b\rangle$, and is shifted by TI1.  For the case of TI2, with $\langle b\rangle=0$, the variance of the distribution is given by $\langle b^2\rangle$.  Again for TI2, as one raises $X_{DR}$ upwards through the critical point, the Gaussian distribution spreads suddenly to describe the initiation of bifurcation.  

For convenience, the symbols $\langle \rangle$ for ensemble averaging will be dropped in the discussion below, and it is understood that $b$ and $u$ actually represent $\langle b\rangle$ and $\langle u\rangle$.

In principle, third and higher orders of induction would be dealt with in a similar way as TI1 and TI2.  There may be systems with both $\alpha=0$ and $\gamma=0$ which would require treatments at higher order.  These systems are expected to be rare, however, and given the success demonstrated in Sec.~\ref{sec:more} below in using TI2 on real systems, leaving higher order TI for future work is justified for now.

\subsection{Generalized Entropy with Focus on the Gate}

At this point, I move from a principle that maximizes the entropy production of the entire system to a principle that focuses on the gate.  As I will show, the DR still plays an important role, but only in supplying parameters for the theory focused on describing the gate.  I will treat first and second order cases separately before combining them.  The procedure merely amounts to multiplying $\Phi_1$ and $\Phi_2$ by constants, thus preserving their maximization properties.  These multiplicative constants will be chosen in such a way as to produce the entropy when the DR is restored to equilibrium.

%That the dissipation function has the same dimensions as power, suggests that dividing by temperature and multiplying by a timescale, giving a quantity withe same units as entropy, might result in a thermodynamic function suitable for the study of nonequilibrium systems. 

\subsubsection{First Order}

Multiplying $\Phi_1$ by a term independent of $b$ also gives a potential that is maximized when the gate is stationary.  
%For a couple of reasons it makes sense to 
I define a new thermodynamic potential as  
\begin{equation}
\Psi_1\equiv \left(\frac{\tau^*}{2}\right)\left(\frac{X_{DR}^2-\langle X_{DR}^2\rangle_0}{X_{DR}^2}\right)\Phi_1=\left(\frac{\tau^*}{2}\right)\left(\frac{X_{DR}^2-\langle X_{DR}^2\rangle_0}{X_{DR}^2}\right)\sigma_{DR}-\frac{\tau_{GT}}{2}\sigma_{GT}~\, . \label{Psi1}
\end{equation}
Though $\Psi_1$ has the same units as entropy, it is not the same as the total system entropy.  It is a nonequilibrium potential since it depends on $X_{DR}$.  In the limit where $X_{DR}^2$ is small, and approaches $\langle X_{DR}^2\rangle_0$, $\Psi_1$ takes the limit of $-\frac{1}{2}\tau_{GT}\sigma_{GT}=-\frac{1}{2}g_{GT}b^2$ which is equal to the difference $\Delta S_{GT}=S_{GT}-S_{GT,eq}$ when this difference is taken up to second order in $b$.  At, or near, $X_{DR}=0$, the guiding principle of maximizing $\Psi_1$ for determining $b$, is to maximize $S_{GT}$, which of course is the governing thermodynamic principle for the gate at, or near, equilibrium.  Thus, maximizing $\Psi_1$ has the attractive feature of determining the (stationary) state of the gate when the DR is driven well away from equilibrium and also seamlessly determining the equilibrium state of the gate when the DR is at equilibrium.  

The potential function $\Psi_1$ is an excellent tool for determining nonequilibrium thermodynamics specifically for the gate.  The DR is effectively separated away from the gate, much like how in equilibrium thermodynamics, the thermal reservoir (heat bath) is mathematically separated away from the system of interest to produce the Helmholtz free energy of that system.
With the traditional heat bath the only process that matters for the system of interest is the transfer of thermal energy with the bath.  Transferring thermal energy will increase the entropy of one of either that system or the bath, at the expense of the other; There is then one energy value that maximizes the total entropy.  The rate at which the bath increases entropy with energy is described in a simple manner by one parameter, $\beta=1/k_B T$.

Here, at a fixed $X_{DR}$, adjusting $b$ will increase one of $\sigma_{DR}$ or $\sigma_{GT}$ at the expense of the other (while the gate is stationary).  One particular value $b_{ss}$ will maximize $\Psi_1$.
From Eq.~(\ref{Psi1}), $\Psi_1$ may be rewritten as:
\begin{equation}
\Psi_{1}(b)=\Psi_{1,0}+k_B\beta_1 b -\frac{1}{2}g_{GT}b^2    ~\, , \label{psi1b}
\end{equation}
where $\Psi_{1,0}=\frac{1}{2}L_{DR}X_{DR}^2\tau^*$, $\beta_1=\gamma X_{DR}^2\tau^*/k_B$, and with fluctuations in $X_{DR}$ ignored. 
 The rate at which the DR increases entropy production with $b$ is described by one parameter, $\beta_1$.  Theoretically, $\beta_1$ for the DR is analogous to $\beta$ for the heat bath, or the chemical potential $\mu$ for a particle bath.  Since $\beta_1$ vanishes when $X_{DR}=0$ it is not surprising that this parameter was not discovered during investigations of equilibrium thermodynamics. 
%For TI2 it would be $\alpha$, not $\gamma$.  
Because of the similarities with how the Helmholtz free energy is formulated to account for interactions of a system with a heat bath, the term \textit{free entropy} is apt for $\Psi_1$.  One thinks then of some amount of entropy as being available in some sense; In particular the gate (dynamically) lends entropy to the dynamical reservoir.
 
%-not sure about this:
%In principle $\gamma$  could be varied in the same way temperature or chemical potential are controlled, and one could study, for example, how systems have phase transitions as this parameter is varied.  One could also do the same by varying $X_{DR}$ at fixed $\gamma$.   For equilibrium thermodynamic systems there is no analogue to the variable $X_{DR}$.
%While there is no adjustable parameter like $X_{DR}$ in equilibrium systems, one can alter the bath temperature, and move through phase transitions this way, (or alter the chemical potential with a particle bath).  The situation is both similar and different with nonequilibrium systems.  One can alter $X_{DR}$ easily enough, but controlling $\gamma$ is not likely to be easy in practice.
%
%This even though the direct analogy is between temperature and $\gamma$, one is more likely to make a loose association between XDR and temperature.
%
%Having XDR variable indicates the possibility for nonequilibrium science being much richer than equilibrium thermodynamics.  For equilibrium there is only one value for $X_{DR}$, namely zero, while in nonequilibrium 

\subsubsection{Second Order}

In terms of $\Delta b^2$, the thermodynamic potential $\Psi_2\equiv\frac{\tau^*}{\mu}\Phi_2$ becomes:
\begin{equation}
\Psi_{2}(u)=\Psi_{2,0}+k_B\beta_2 u -\frac{g_{GT}}{8Q^2\langle b^2\rangle_0}u^2    ~\, , \label{psi2b}
\end{equation}
where $\Psi_{2,0}=L_{DR}X_{DR}^2\tau^*/\mu+\frac{1}{2}g_{GT} z^2 \langle b^2\rangle_0$, and $\beta_2=g_{GT}h_+/2k_B$. 
%\begin{equation}
%\Psi_2(u)=L_{DR}X_{c}^2\tau^* \frac{z^2}{\mu}+\frac{g_{GT}}{2}\left[ z^2\langle b^2\rangle_0 +\frac{z^2}{\mu} u +(z^2-1)u-\frac{1}{4Q^2\langle b^2\rangle_0}u^2\right] ~\, . \label{Psi2mod}
%\end{equation}
%\begin{equation}
%\Psi_2(u)=L_{DR}X_{DR}^2\tau^* \frac{1}{\mu}+\frac{g_{GT}}{2}\left[ z^2\langle b^2\rangle_0 + u h_+(z^2)-\frac{u^2}{4Q^2\langle b^2\rangle_0}\right] ~\, . \label{Psi2mod}
%\end{equation}
This free entropy function is maximized both above and below critical when $u=u_{ss}$, and the second derivative at $u=u_{ss}$ is always negative.  Near the maxima, well above critical, the dependence for $\Psi_2$ near $b=\pm b_{ss}$ goes as $-g_{GT}z^2(b\pm b_{ss})^2$ which is on the same order as the dependence of $S_{GT}-S_{GT,ss}=-\frac{1}{2}g_{GT}b^2$ at equilibrium.
Well below critical, terms in $\Psi_2$ that go as $u$ disappear, so the potential flattens out, and the induction essentially disappears. 
Also,
%\begin{equation}
%\lim_{X_{DR}\to 0}\Psi_{2}(u)=\Psi_{0}-\frac{c^2 u^2}{2L_{GT}} =\Psi_{0}-\frac{1}{2}g_4 u^2  ~\, , \label{Psi2lim}
%\end{equation}
\begin{equation}
\lim_{X_{DR}\to 0}\Psi_{2}(u) =\Psi_{0}-\frac{1}{2}g_4 u^2  ~\, , \label{Psi2lim}
\end{equation}
where $\Psi_{0}=L_{DR}X_{c}^2\tau^*$.  
%In the expression on the far right of Eq.~(\ref{Psi2lim}) I have made use of entropy-based form of the cutoff discussed in Sec.~\ref{sec:SOTI} where $c^2=g_4 L_{GT}$.

Given the definition of the second order variable $u$, with inspection of Eq.~(\ref{g2g4}), the best expression for describing small changes from equilibrium of the gate entropy, up to second order, is:
\begin{equation}
\Delta S_{GT}=S_{GT}-S_{GT,eq}=-\frac{1}{2}g_{2}b^2-\frac{1}{2}g_{4}u^2~\, .  \label{SIentbu}
\end{equation}
Equation~(\ref{SIentbu}) allows for the connection of the first and second order potentials, simply by adding $\Psi_1$ and $\Psi_2$.  This is consistent with the $z=0$ limit:
\begin{equation}
\lim_{X_{DR}\to 0}\left(\Psi_{1}(b)+\Psi_{2}(u)\right)=\Psi_{0}+S_{GT}-S_{GT,eq} ~\, . \label{Psi12lim}
\end{equation}

\subsubsection{Combining up to Second Order} \label{uptp2}

Establishment of both $\Psi_1$ and $\Psi_2$ places the PMEP on a solid footing for a great many physical systems; For those thermodynamic systems where the gate symmetry is already broken in equilibrium, one would use and maximize $\Psi_1$, and for systems where symmetry in thermodynamic equilibrium is not broken, $\Psi_2$ is the potential chosen to be maximized.  
If one forms a potential function $\Psi\equiv \Psi_1(\langle b\rangle)+\Psi_2(\Delta b^2)$ as the sum of first and second order potentials then one must consider the free entropy $\Psi$ as a function of two variables, the mean, and the dispersion of $b$ and maximize with respect to each variable separately.
%Due to the all or nothing nature of 2nd order TI, it is plausible that both a dynamical equation and a function $\Psi$ can be formulated that account for all TI up to 2nd order.  For $\Psi$, this amounts to compiling $\Psi_1$ and $\Psi_2$ while properly accounting for the non-TI terms such as $L_{DR}X_{DR}^2$ and $g_{GT}b^2$.  Below critical, 1st order TI dominates, while at and above critical, 1st order TI would supply only small corrections.  
In practice, one may very well use only one of the $\Psi_1$ and $\Psi_2$ potentials in a given problem. This usage would be determined by symmetry considerations.
%The physics of the SI dictate that the product $\alpha\gamma=0$, i.e., one or both of the parameters $\alpha$ and $\gamma$ are zero, dictated by symmetry considerations.

%Because $\Psi$ has the same dimensions as entropy and because of the interaction of the SI with the DR and the idea that the SI borrows entropy from the DR, I believe the term \textit{free entropy} is an apt one for $\Psi$.

Since $\Psi$ is a linear combination of $\Phi_1$ and $\Phi_2$, it satisfies the stationary state PMEP with regards to both $b$ and $u$.  This alone is a key result and allows for the formulation of variational principles when order parameter fields are discussed below.  The specific linear combination that I have chosen will connect nicely to the traditional entropy when the DR is in equilibrium.
Equation~(\ref{Psi12lim}) shows that, apart from the constant term $\Psi_0$, $\Psi_1+\Psi_2$ approaches $S_{GT}-S_{GT,eq}$ as $X_{DR}$ approaches zero.  Thus the quantity $S_{GT,eq}+\Psi-\Psi_0$ is the gate entropy at or near equilibrium and would be maximized according to equilibrium thermodynamics.  Well away from equilibrium, I have found that $S_{GT,eq}+\Psi$ is also maximized, not by the equilibrium state, but by the stationary state.  Of course, the stationary state becomes the equilibrium state as $X_{DR}$ is turned down to zero.  Thus, the potential, 
\begin{equation}
\mathpzc{S}_{NE}\equiv S_{GT,eq}+\Psi-\Psi_0  \,,  \label{SNE}
\end{equation}
differs from $\Psi$ by only constants and makes a seamless transition to the entropy as $X_{DR}$ goes to zero, i.e.,  
\begin{equation}
\lim_{X_{DR}\to 0}\mathpzc{S}_{NE}=S_{GT}   ~\, . \label{3rdlaw}
\end{equation}
This new potential, $\mathpzc{S}_{NE}$, is not the entropy; The standard entropy function provides an incomplete description well away from equilibrium. The potential $\mathpzc{S}_{NE}$ is what the relevant thermodynamic potential becomes when the DR is well away from equilibrium, and $\mathpzc{S}_{NE}$ can be thought of as the \textit{generalized entropy}. 
%  A generalized form of entropy for nonequilibrium study is not merely of the form $-gb^2$.

Equivalently, one may define the \textit{excess entropy}, $\Xi\equiv\Psi-\Psi_0+\frac{1}{2}g_{2}b^2+\frac{1}{2}g_{4}u^2$, which leaves the generalized entropy expressed as
\begin{equation}
\mathpzc{S}_{NE}\equiv S_{GT}+\Xi  \,.  \label{SNE2}
\end{equation}
This bookkeeping allows one to identify $\Xi$ as something completely outside of the realm of equilibrium thermodynamics, i.e., $\Xi=0$ when $X_{DR}=0$ even if some other agent pushes $b$ away from zero.
%The function $\Xi$ also summarizes nicely my results presented here.
Explicitly then,
%\begin{equation}
%\Xi(b,u)=\frac{1}{2}L_{DR}X_{DR}^2\tau^*+L_{DR}X_{DR}^2\tau^*\left(\frac{1}{\mu}-\frac{1}{z^2}\right)+\frac{1}{2}g_{GT} z^2\langle b^2\rangle_0 +k_B\beta_1 b+k_B\beta_2u \,.  \label{Xi}
%\end{equation}
\begin{equation}
\Xi(b,u)= \Xi_0+k_B\beta_1 b+k_B\beta_2u \,,  \label{Xi}
\end{equation}
where $\Xi_0=L_{DR}X_{DR}^2\tau^*\left(\frac{1}{2}+\frac{1}{\mu}-\frac{1}{z^2}\right)+\frac{1}{2}g_{GT} z^2\langle b^2\rangle_0$.
Since each term in Eq.~(\ref{Xi}) contains a factor of $\tau^*$, $\Xi$ is essentially an entropy created during the time span of a scattering event so it may be referred to as a \textit{dynamical entropy}.  The term \textit{rheological entropy}  may also be apt.
It's also important to note the additive nature of Eq.~(\ref{SNE2}); This is not a trivial result, as there are many ways to mathematically extend $S$ to the nonequilibrium realm ($S$ being shortform for $S_{GT}$).

At this point it is helpful to summarize physical laws as a three-tier structure: (1) standard mechanics, both classical and quantum, taking place at zero temperature where $S=0$ (assuming no degeneracy in the ground state) and $\Xi=0$, (2) equilibrium thermodynamics where the temperature is raised above zero, $S\neq 0$, and $\Xi=0$, and (3) nonequilibrium thermodynamics where $X_{DR}$ is adjusted from zero,  $S\neq 0$, and $\Xi\neq 0$.
In tier 1, the guiding principle is finding extrema in the Lagrangian. In tier 2, the guiding principle is maximizing $S$, and in tier 3, the guiding principle is maximizing $\mathpzc{S}_{NE}$.  
%Inside of this 3-tier structure are nonequilibrium analogs to both the second and third laws of (equilibrium) thermodynamics. 

This three-tier structure bears a strong resemblance to the laws of thermodynamics, especially how maximizing a generalized entropy $\mathpzc{S}_{NE}$ is so similar to the SLT, i.e., maximizing entropy in equilibrium thermodynamics.  Also, having $\Xi=0$ when $X_{DR}=0$ bears strong resemblance to having $S=0$ when $T=0$.  
%Though formulating new physical laws is no small task, 
These results are compelling and lead to the following principles for governing nonequilibrium systems: 

\vspace{0.2in}
Nonequilibrium Principle 1: The generalized entropy for the gate is additive:  $\mathpzc{S}_{NE}=S+\Xi$,

\vspace{0.2in}
Nonequilibrium Principle 2: $\mathpzc{S}_{NE}$ is maximized when the gate is in a stationary state,

\vspace{0.2in}
Nonequilibrium Principle 3: $\Xi=0$ when the DR is in equilibrium.
\vspace{0.2in}

It's important to point out that the gate is not the entire system;  For the entire system the guiding principle is still the SLT, i.e., the system always evolves so that $S_T$ is eventually maximized.  What these three new principles do is describe key aspects of the dynamics as the entire system evolves towards equilibrium.
If these principles are in time to become established as laws there must always exist a function $\Xi$ that satisfies these three laws, no matter how complicated the system.  This function exists and the three laws work for a pitchfork bifurcation.  The question remains about whether they also apply to more complicated instabilities, ex. Hopf and Takens-Bogdanov bifurcations~\cite{Pearson1989,Castets1990}.  

It also remains to be seen if these results apply to systems that are not purely dissipative, i.e., for systems that have inertial degrees of freedom as well as damping.
Some discussion of the physical significance of the variable $b$ in Eqs.~(\ref{aqdot1stcorr}) and (\ref{bdot}) is warranted.  If $b$ is a mechanical variable then under certain circumstances the dynamics can be purely dissipative, for example the LR circuit and mechanical equivalent, an otherwise free particle with damping.  If a small spring constant is added, the system is highly overdamped and is still well-approximated by purely dissipative dynamics.  However, when the damped harmonic oscillator becomes underdamped, it becomes difficult to define a stationary state in the pure sense ($\dot{b}=0$) because of the natural oscillations. 
In this case the dynamics described in Eq.~(\ref{bdot}) would be for the envelope of $b(t)$.  This approach is quite effective for the highly underdamped case where the oscillation period is much less than the relaxation time.
This, essentially, is coarse-graining in time, and it is understood that some detailed information about the system is filtered out of the dynamical equations.

A fundamental assumption for thermodynamic variables is that they provide an incomplete description;  Describing a macroscopic system with only several variables means that there are many distinct microstates corresponding to a given thermodynamic state, in this case for a given value of $b$.  It is reasonable then that after coarse graining, the dynamics of thermodynamic variables would be purely dissipative even when the dynamics for internal microscopic dynamics is not.  Stationary states would then always be achievable and the principle of maximizing $\mathpzc{S}_{NE}$ will always be valid. 

Given the paucity of  general thermodynamic results in the nonequilibrium realm~\cite{Nicolis1977}, I anticipate these laws as building on the Onsager symmetry relations to constitute the current extent of such general knowledge.  
Indeed the Onsager symmetry relations might be referred to as the initial (or perhaps zeroth) law of nonequilibrium thermodynamics.

%Below critical, $\lambda$ is indeterminate since the stationary state is $b=0$. Physically, one would expect $\Phi$ to be continuous through the critical point with respect to $X_{DR}$, and then should stiffen below critical.  If $f(x)=x^2-1$ for $x<1$, and $f(x)=1-1/x^2$ for $x>1$, then I propose that below critical $\Psi=f(X_{DR}/X_c)\tau^*\sigma_{DR}/2-\tau_{GT}\sigma_{GT}/2$ such that for all $X_{DR}$:
%\begin{equation}
%\Psi= \frac{\tau^*}{2} f(X_{DR}/X_c)\sigma_{DR}-\frac{\tau_{GT}}{2}\sigma_{GT}   ~\, . \label{Psi}
%\end{equation} 
%This makes $\Psi$ continuous through the phase transition, finite at all values of $X_{DR}$.  Another benefit from defining $\Psi$ in this manner concerns the limit as $X_{DR}$ approaches zero.  When $X_{DR}$ is set to zero, $\sigma_{DR}$ disappears, as well as the induction term in $\sigma_{GT}$, leaving only the dissipative term for the gate.  For small $b$, $\Psi=-g_{GT}b^2/2$ and this is precisely the gate entropy change relative to the equilibrium value $S_{GT,eq}$.  Thus,
%display mode looks good!
%\[ \lim_{x \to 2} f(x) = 5 \]
%Discounting degeneracies, S goes to zero as T goes to zero. is similar to Psi to zero as XDR to zero.

\subsection{Nonequilibrium Le Chatelier's Principle}

%The exponential growth after bifurcation, towards the stationary state, suggests that second order TI could create these structures faster and with greater strength than with first order TI alone, especially if $c$ is small. 

In this section I establish a nonequilibrium version of Le Chatelier's principle, for the case of second order TI.  For TI1 this principle has already been established~\cite{Patitsas2014} and this treatment follows my previous one closely.

If the DR is pushed away from equilibrium, while the gate is held fixed in its equilibrium state, then the total rate of entropy production $\sigma_T$ coincides with $\sigma_{DR}|_{\{X_{GT}=0\}}=L_{DR}X_{DR}^2$.  The question then is in determining how this rate compares with that when the gate is allowed to relax.  If the gate is released at $t=0$ then right after $t=0$, as $u$ grows from zero, the gate would appear to violate the SLT if an observer was unaware of the DR, since $\sigma_{GT}=-g_4 u\dot{u}$.  By Eq.~(\ref{b2mod2})
\begin{equation}
\sigma_{GT}=-g_4 u\left[4\alpha L_{GT}\tau^*\left(X_{DR}^2-X_c^2\right)u+4\alpha L_{GT}\tau^* X_{DR}^2\langle b^2\rangle_0 -4c^2 u^2\right] ~\,.   \label{sigGT}
\end{equation}
As $u=0$ at $t=0$, for small $t$,  $u= 4\alpha L_{GT}\tau^* X_{DR}^2\langle b^2\rangle_0 t$ to leading order, and $\sigma_{GT}=-\frac{\tau^*}{\tau_{GT}Q^2} \alpha X_{DR}^2 u$. In the meantime $\sigma_{DR}=L_{DR}X_{DR}^2+\alpha X_{DR}^2 u$ and the net rate is $\sigma_{T}=L_{DR}X_{DR}^2+\alpha X_{DR}^2 u (1-\frac{\tau^*}{\tau_{GT}Q^2})$.  The relaxation time $\tau_{GT}$ is always larger than $\tau^*$ which is typically the fastest time scale in the analysis.  Also, $Q>>1$ typically, so there is no risk of $\sigma_T$ being negative for short times.  Indeed, the relaxation of the gate increases $\sigma_T$.

For longer times, $u^2$ becomes significant, but only if $z\geq 1$.  The rate $\sigma_{GT}$ increases for some time before it eventually decreases and falls to zero when the gate becomes stationary.  
The maximum for $\sigma_{GT}$ occurs at $u_{max}=2\langle b^2\rangle_0 (z^2-1)$ at which  $\sigma_{GT}=-(\tau^*/\tau_{GT})\alpha X_{DR}^2 u_{max}[(z-1/z)^2+1/Q^2]$.  With $Q$ normally much greater than unity and $\sigma_{GT}$ does not overcome the $\alpha X_{DR}^2 u$ term in $\sigma_{DR}$, unless $z^2$ is very large, i.e., on the order of $\tau_{GT}/\tau^*$.  
Achieving $z$ values of 1 or 2 is not easy in real systems, and making $z^2$ on the order of $\tau_{GT}/\tau^*$ could lead to catastrophic conditions for real systems.  Also, even if $z^2$ could get this large there is still the $L_{DR}X_{DR}^2$ term (from $\sigma_{DR}$) to overcome, making violation of the SLT unlikely.  Nevertheless it is interesting to note that under conditions with very large $X_{DR}$ there is a possibility of briefly violating the SLT when the gate is a very small system with a relaxation time approaching $\tau^*$.

%
%Right after bifurcation, with $X_{DR}^2>X_c^2$, while approaching the stationary state,  $\sigma_{GT}=-\frac{\tau^*}{\tau_{GT}} L_{GT}( X_{DR}^2-X_{c}^2)<0$, and the gate would appear to violate the SLT, if an observer was unaware of the DR.  With the DR accounted for,  $\sigma_{DR}=L_{DR}X_{DR}^2+2\alpha b^2 X_{DR}^2$,  the total system entropy production, right after bifurcation, $\sigma_T=\sigma_{DR}+\sigma_{GT}=X_{DR}^2[L_{DR}+\alpha b^2(1-2\tau^*/\tau_{GT}(1-X_c^2/X_{DR}^2))]$.  The non-negativity of $\sigma_T$ is guaranteed when $\alpha>0$ and $\tau^*/\tau_{GT}<0.5$.  
%In a similar way one can show that the same conclusions hold for later times, even when $b$ is increasing exponentially.

For even later times, the gate settles into its stationary state where $\sigma_{GT}=0$ and the positive definite nature of $\sigma_{DR}$ guarantees the total entropy always increases. More careful inspection reveals an even stronger result:
% In fact, TI increases the total rate of entropy production from what it is otherwise.  
%The situation is much the same when the gate is in its stationary state.
%and in the stationary state:
%\begin{equation}
%\sigma_{T,ss}=L_{DR}X_{DR}^2 +\frac{\alpha^2}{c} L_{GT}\tau^* X_{DR}^2 ( X_{DR}^2-X_{c}^2)  ]~\, . \label{sigmaSS}
%\end{equation}
If $\sigma_{DR} |_{\{X_{GT}=0\}}$  is the total entropy production with the gate at equilibrium, and $\sigma_{DR} |_{\{J_{GT}=0\}}$  is the total entropy production with the gate stationary (so the flux $J_{GT}=\dot{b}$ is zero) then one always has the condition   
\begin{equation}
\sigma_{DR} |_{\{J_{GT}=0\}} \geq \sigma_{DR} |_{\{X_{GT}=0\}} ~\,. \,\,\,\,\,\,\,\,\,\,\,\,\text{(2nd order noneq. Le Chatelier's principle)}  \label{LeChat}      
\end{equation}
The same result was previously established for TI1~\cite{Patitsas2014}. 

Similarly to TI1, the physical meaning of this result is that the DR produces entropy faster when the gate variable is allowed to relax by bifurcating away from equilibrium and becoming stationary.  This result constitutes a nonequilibrium version of Le Chatelier's principle.  In the traditional Le Chatelier's principle, when a given thermodynamic variable is pushed away from equilibrium, other thermodynamic variables relax to new equilibrium values, so that the total entropy is again maximized, and the new relaxed entropy is always greater than the unrelaxed entropy~\cite{LandauL}.  Here, when the DR variable is pushed away from equilibrium, the gate variable will temporarily move away from equilibrium to stationary states, and the new rate of entropy production is always greater.  The actual difference is 
%\begin{equation}
%\sigma_{DR} |_{\{J_{GT}=0\}}- \sigma_{DR} |_{\{X_{GT}=0\}}= \frac{2\alpha^2\tau^* L_{GT}}{c^2}X_{DR}^2 ( X_{DR}^2-X_{c}^2)   ~\,,   \label{LeChatdiff}      
%\end{equation}
\begin{equation}
\sigma_{DR} |_{\{J_{GT}=0\}}- \sigma_{DR} |_{\{X_{GT}=0\}}= \frac{\tau^* L_{GT}}{c^2}\alpha^2 X_{DR}^4 \left(\frac{h_+}{z^2}\right)   ~\,.   \label{LeChatdiff}      
\end{equation}
The right hand side of Eq.~(\ref{LeChatdiff}) is always positive definite.  Since the entropy production rate $\sigma_{GT}$ of the gate is zero when the gate is stationary, the nonequilibrium Le Chatelier's principle also applies to the total rate of entropy production, $\sigma_T=\sigma_{DR}+\sigma_{GT}$, then the principle works also for $\sigma_T$, i.e., $\sigma_{T} |_{\{J_{GT}=0\}} \geq \sigma_{T} |_{\{X_{GT}=0\}}$.  In fact, the principle will work for any quantity formed by the sum of $\sigma_{DR}$ and any multiple of $\sigma_{GT}$.
%\subsubsection{Entropy Coupling Problem}
%
%The nonequilibrium Le Chatelier's principle up to 2nd order along with Eq.~(\label{Sabove}) for the entropy budget
%While the nonequilibrium Le Chatelier's principle is not a proof of the existence of a guiding extremal principle, it does suggest the possibility.
The induction effect always causes the DR to approach equilibrium faster, thus leading to the conclusion that the gate will always \textit{facilitate} the DR's approach to equilibrium.  

% is motivation for a general principle which maximizes the rate of approach to equilibrium in some manner.

\subsection{Entropic Coupling Problem} \label{Sec:EntCoup}

In the second order facilitation of the approach to equilibrium, the gate will be shifted away from its own thermodynamic equilibrium, and if $X_{DR}$ is large enough compared to $X_c$, the disruption of the gate equilibrium may be detected.
The stationary state result Eq.~(\ref{uquad}) suggests that for TI2 the gate entropy change from equilibrium is given by
%\begin{equation}
%\Delta S_{GT,ss}=-2k_B Q^2\left[z^2-1+\sqrt{(z^2-1)^2+z^2 /Q^2}\right]<0    \,.   \label{SGTquad}
%\end{equation}
\begin{equation}
\Delta S_{GT,ss}=-\frac{1}{2}k_B Q^2 h_+^2 <0    \,.   \label{SGTquad}
\end{equation}
For consistency this does coincide at large $z^2$ with the infinite $Q$ result using the scheme $g_{GT}=g_2+g_4 b^2$.
This change is important regarding the entropic coupling problem, which will be further discussed in Sec.~\ref{Sec:EntCoup}. 
One verifies that the entropy is continuous through the transition at $z=1$ and therefore the physics described here is similar to that of a second order (equilibrium) PT.  Equation~(\ref{SGTquad}) provides a sort of budget that is available for forming patterned structures in the gate.

Equation~(\ref{SGTquad}), along with the corresponding first order result discussed in Ref.~\cite{Patitsas2014} and the nonequilibrium Le Chatelier's principle, up to second order, solves the entropic coupling problem.  The gate can achieve a state that may be considered to be patterned, or self-organized/assembled, by having its entropy lowered relative to its equilibrium, and during the entire time of this self-assembly, the SLT is never violated.  The result holds for both TI1 and for TI2, both below and above the critical bifurcation point.  The DR plays a key role as that system which always increases it entropy by at least as much as the decrease in the entropy of the gate.  I stress the dynamical nature; If the DR returns to equilibrium, the entropy budget $\Delta S_{GT,ss}$ goes to zero. 

This type of coupling between the DR and the gate is unconventional.  Ordinarily one couples two systems mechanically via a potential energy term.  When the potential energy depends on variables from the two systems then the dynamical differential equations become coupled.  Here, the physics is purely dissipative and potential and kinetic energies do not play a role.  Instead, it is the entropy production that depends on both DR and gate variables, and produces the interesting couplings in the dynamics.  More specifically, the ultimate source of the coupling is that the DR conductance coefficient $M_{DR}$ depends on the gate variable.  One can say that the conductance becomes paramount in nonequilibrium systems, playing a role similar to the potential energy in equilibrium systems.  

I will resume discussion on this problem below in Sec.~\ref{Sec:EntCoup2}.

\subsection{Nonequilibrium Canonical Distribution} \label{sec:canon}

For (isolated) systems in equilibrium, all accessible microstates receive equal statistical weight, according to the postulate of equal a priori probabilities~\cite{Reif,Tolman79}.  In this section I show how this changes in nonequilibrium and I derive revised weighting functions.

The thermodynamic potentials $\Psi_1$ and $\Psi_2$ may be used to calculate nonequilibrium partition functions for the gate specifically.  These would be multiplied by the regular equilibrium partition functions to provide the full picture.  I define the first order partition function as
\begin{equation}
Z_{1,noneq}\equiv c' \int_{-\infty}^{\infty}e^{\Psi_1(b)/k_B}db    ~\, , \label{Z1a}
\end{equation}
where $c'$ is a constant to be determined below.  Using Eq.~(\ref{psi1b}) the integral is easily evaluated as:
\begin{equation}
Z_{1,noneq}= c'~ e^{\Psi_{1,0}/k_B}\sqrt{\frac{2\pi k_B}{g_{GT}}}e^{k_B\beta_1^2/2g_{GT}}    ~\, . \label{Z1b}
\end{equation}

The same result for $Z_{1,noneq}$ can be found by focusing on microstates of the gate and it is instructive to do this analysis.
The variable $b$ represents a macrostate to which there are $\Omega_{GT}(b)$ gate microstates all corresponding to the same $b$.  The probability of the gate having value $b$ is proportional to the total multiplicity: $\Omega_T=\Omega_{DR}\Omega_{GT}(b)$.  The assumption is that all microstates for the \textit{entire} system are accessed equally.
The DR multiplicity $\Omega_{DR}$ does not depend directly on $b$, but its rate of change does because the Onsager coefficient $M_{DR}$ depends on $b$.
The way to physically interpret $\Psi$ is to not evaluate $\Omega_{DR}$ at time $t$ but rather at $t+\tau^*/2$, i.e., to reach forward in time the time required for a scattering event. This means using not $\Omega_{DR}$ but $\Omega_{DR}\exp{(\sigma_{DR}\tau^*/2k_B)}$ when attempting to derive a probability distribution.
%This situation is specially suited for the potential $\Psi_1$ since careful use of $\Psi_2$ associates to the variable $u$, not $b$. 
The probability of the gate having value $b$ is proportional to $\exp{(\Psi_1(b)/k_B)}$.
Noting that 
\begin{equation}
\Omega_{GT}(b)=\Omega_{GT,eq}\exp{(-g_{GT}b^2/2k_B)} ~, \label{OmGT}
\end{equation}
which is independent of any TI effects, and inspecting Eq.~(\ref{psi1b}), the probability of the GT having value $b$ is found to be proportional to $\exp{(\beta_1 b)}\Omega_{GT}(b)$.
%, where $\beta_1=\tau^* X_{DR}^2 \gamma/k_B$.

In equilibrium, the probability of any particular gate microstate being occupied is equal to any other.  Away from equilibrium, this is not the case;
For each gate microstate $i$, with corresponding $b_i$, the probability of occupancy is the \textit{nonequilibrium canonical distribution}, 
\begin{equation}
P_{1,i}=c_1 e^{\beta_1 b_i}   ~\, . \label{canon}
\end{equation}
%where $\beta_1=\tau^* X_{DR}^2 \gamma/k_B$.
The normalization constant $c_1$ is evaluated in the standard manner, invoking a partition function: 
\begin{equation}
c_1^{-1}=Z_{1,noneq}=\sum_i e^{\beta_1 b_i} =\int_{-\infty}^{\infty}{db \, e^{\beta_1 b_i}\Omega_{GT}(b)}   ~\, . \label{part}
\end{equation}
Inserting $\Omega_{GT}(b)$ from Eq.~(\ref{OmGT}) into Eq.~(\ref{part}), allows the integral to be evaluated, leading to:
%Since $\Omega_{GT}(b)=\Omega_{GT,eq}\exp{(-g_{GT}b^2/2k_B)}$, 
\begin{equation}
Z_{1,noneq}=\Omega_{GT,eq}\sqrt{2\pi k_B/g_{GT}}\exp{(k_B\beta_1^2 /2g_{GT})}  \, . \label{Z1}
\end{equation}
After comparing to Eq.~(\ref{Z1b}), one identifies the constant $c'=\Omega_{GT,eq}e^{-\Psi_{1,0}/k_B}$.

From Eq.~(\ref{Z1}) one readily verifies that $\langle b\rangle=\partial\ln{Z_{1,noneq}}/\partial\beta_1$, which equals the expected value $b_{ss}$.  Also $\langle (\Delta b)^2\rangle=\partial^2\ln{Z_{1,noneq}}/\partial\beta_1^2 =k_B/g_{GT}$.  %Further from equilibrium, $g_{GT}$ might vary with $b$ and Eq.~(\ref{Z1}) would require corrections. 
%I expect that Eq.~(\ref{part}) would have very general applicability, for all forms of  $\Omega_{SI}(b)$.
From Eq.~(\ref{part}) one notes that to a good approximation:
%\begin{equation}
%k_B\ln Z_{1,noneq}= k_B\ln \Omega_{SI,eq}+k_B\beta_1 b -\frac{1}{2}g_{GT}\langle b\rangle^2  \, .
%\end{equation}
\begin{equation}
k_B\ln Z_{1,noneq}= S_{GT,eq}+k_B\beta_1 \langle b\rangle -g_{GT}\langle b\rangle^2/2  \, ,
\end{equation}
which can be expressed as $k_B\ln Z_{1,noneq}=S_{GT}+k_B\beta_1 \langle b\rangle $.

Using Eq.~(\ref{psi2b}), the second order partition function is then
\begin{equation}
Z_{2,noneq}\equiv \int_{-\infty}^{\infty}e^{\Psi_2(u)/k_B}du= e^{\Psi_{2,0}/k_B}\sqrt{\frac{8\pi Q^2\langle b^2\rangle_0}{g_{GT}}}\exp{\left(\frac{2k_B Q^2\langle b^2\rangle_0\beta_2^2}{g_{GT}}\right)}    ~\, . \label{Z2a}
\end{equation}
To a good approximation $Z_{2,noneq}\approx e^{\Psi_2(\langle u\rangle)/k_B}$.

In order to obtain the probability distribution for gate states, $i$, with values $u_i$ one notes that  the multiplicity function for the gate is a function of both $b$ and $u$: $\Omega_{GT}(b,u)=\Omega_{GT,eq}\exp{(-g_{2}b^2/2k_B-g_{4}u^2/2k_B)}$.  Proceeding similarly to the first order case the gives:
\begin{equation}
P_{2,i}=\frac{1}{Z_{2,noneq}} e^{\beta_2 u_i}   ~\, . \label{canon2}
\end{equation}
For $z<<1$ this probability distribution is flat, coinciding with equilibrium expectations.  Near and above the critical point at $z=1$ the situation changes and larger values of $\Delta b^2$ are favoured, leading to the bifurcation. 
Combining first and second order for a total nonequilibrium partition function gives $Z_{noneq}=Z_{1,noneq}Z_{2,noneq}=\Omega_{GT}\exp(\beta_1 \langle b\rangle+\beta_2 \langle u\rangle)$. 

In equilibrium thermodynamics the partition function is simply related to the Helmholtz free energy. 
The analysis presented so far in this section leads naturally to the definition of a generalized free energy as
%When Zeq is added in 
%
%$k_B\ln Z_{eq} Z_{1,noneq}\approx S_{SI}+k_B\beta_1 b -U/T$
%
%$-k_B T\ln Z_{eq} Z_{1,noneq}\approx U-TS_{SI}-k_B T\beta_1 b =F-k_B T\beta_1 b$
\begin{equation}
\mathpzc{F}\equiv F-k_B T\beta_1 \langle b\rangle-k_B T\beta_2 \langle u\rangle \, ,
\end{equation}
where $F=U-TS$ is the Helmholtz free energy, with $U$ and $S$ being understood as the internal energy and entropy, respectively, of that part of the gate that is removed from the heat bath.  Equation~(\ref{Xi}) can be used to express this generalized free energy as $\mathpzc{F}=F-T(\Xi-\Xi_0)$.
%$-k_B T\ln Z_{eq} Z_{1,noneq}\approx U-TS_{SI,eq}-T\Delta S -k_B T\beta_1 b $
%Apart from a constant offset stemming from the term $\frac{1}{2}L_{DR}X_{DR}^2\tau^*$ in the expression for $\Psi_1$ one may write:
%\begin{equation}
%\mathpzc{F}_1= U-TS_{SI,eq}-T\Psi_1 \, .
%\end{equation}
%This is minimized with respect to $b$, as well as the thermodynamic variables that F depends on.
%Invoking the considerations from Sec.~(\ref{uptp2}) allows one to express $\mathpzc{F}$ in terms of $\mathpzc{S}_{NE}$ as
%\begin{equation}
%\mathpzc{F}\equiv U-TS_{SI}-T\Xi = U-T\mathpzc{S}_{NE} \, . \label{Fnon}
%\end{equation}
%For second order TI, one simply invokes $\exp{(\Psi(b,u)/k_B)}$ and the 2nd order canonical distribution is $P_{2,i}=\exp{(\beta_2 u_i)}$, with $Z_{2,noneq}=\sum_i e^{\beta_2 u_i}$, where $\beta_2=\tau^* X_{DR}^2 \alpha/k_B$. 
%The net partition function for the SI is formed by combining with the equilibrium one as the product $Z_{net}=Z_{eq}Z_{1,noneq}Z_{2,noneq}$ 
%define $\mathpzc{F}\equiv U-T\mathpzc{S}_{NE}$
%Psi1 has SGT in it even if gamma is zero, so yes just add in Psi2(u)
There is a great deal of physics built into the function $\mathpzc{F}$ which has a general, all-encompassing, quality with application to all scientific problems. 
The internal energy $U$ accounts for the mechanics of inter-particle interactions, the entropy $S$ uses the state-counting multiplicity function, often using combinatorial methods, and finally $\Xi$ keeps track of the nonequilibrium fluxes from thermodynamic forces, including TI.

\subsection{Nonequilibrium Phase Transitions}

Well away from equilibrium, $U$ may not have interesting behaviour as $X_{DR}$ changes, even through a nonequilibrium PT.  The internal energy may provide minor corrections, but the key elements of the PT are determined by $\Xi$, not $U$.   
The nonequilibrium PT is clearly distinct from equilibrium PTs at which a free energy $F$ is minimized.  While it is true that at the critical temperature for an equilibrium PT there may be large fluctuations with a nonequilibrium character, this is qualitatively very different from a nonequilibrium PT.  In a nonequilibrium PT, a thermodynamic variable is pushed well away from equilibrium, where $F$ is nowhere near its minimum value and fluctuations from equilibrium would reach only extremely rarely.  When this variable reaches its critical value, the thermodynamic potential $\Psi$ is maximized (and $\mathpzc{F}$ is minimized).  In the above analysis, energetic considerations play no significant role and the free energy $F$ would display no interesting behaviour at the transition.

I also make a note on terminology;  Since the nonequilibrium PT occurs at the bifurcation point and the bifurcation can now be described thermodynamically, then in the context of the thermodynamics of dissipative systems, bifurcation and nonequilibrium PT are interchangeable terms. 

It is instructive to compare nonequilibrium PTs to the second order PTs of equilibrium thermodynamics.  Second order PTs are well described by minimizing the (Helmholtz) free energy function of Landau, $F_L=F_0+\xi_1 b^2 +\xi_2 b^4$, with the transition occurring at $\xi_1 =0$ as $\xi_1$ varies with an external parameter such as temperature.  Below (above) critical, $\xi_1>0$, ($\xi_1<0$), and $F_L$ resembles a parabola (sombrero).
For the nonequilibrium PT well above critical, $\mathpzc{F}$ will resemble the sombrero function because of the form of $\Phi_2$ in Eq.~(\ref{Phi2}).  With $\gamma=0$ (TI2 only), $\Phi_1$ gives the desired parabolic appearance for $\mathpzc{F}$.   The similarity ends there though as for the function $\Phi_2$ to be useful above and below critical it is actually a function of $u$, and $\mathpzc{F}$ is a function of two variables, $\langle b\rangle$ and $u$.  Unlike the equilibrium case, one must minimize  $\mathpzc{F}$ with respect to two variables.  The simple graphical pedagogy used for Landau theory with a varying $\xi_1$ does not work so easily in the nonequilibrium case. 

These nonequilibrium PTs occur when a balance between (gate) entropy and TI is achieved.  In contrast, equilibrium PTs occur when a balance is created between entropy and energy.  Without the energetics, the analysis is of course very different.  Very close to an equilibrium PT it is known that many higher-order terms in the perturbative analysis are required which creates a challenging problem~\cite{Patashinskii}.  While it may be true that a more accurate description of nonequilibrium PTs would require expanding $\exp(\Delta S/k_B)$ to higher than first order, this analysis would differ from the energetic approach that would be described in terms of particle-particle interactions.

Finally, I point out that a second class of nonequilibrium PTs is possible, where a balance is achieved between energy and induction only.  As a possible example, consider a percolation network~\cite{Kirkpatrick1973} that is non-conductive at or near equilibrium, but which becomes conductive well away from equilibrium when the applied bias $X_{DR}$ reaches a certain threshold.  The induced changes in the network that make it conductive may be subtle and purely mechanical.  This indeed may be behind the type of nonequilibrium PTs reported as directed percolation~\cite{HINRICHSEN2006}.

\subsection{Auxiliary Potential for TI2}\label{sec:aux}

The thermodynamic potentials $\Phi_2(u)$ and $\Psi_2(u)$ were devised after the dynamics for $u$ was established in Eq.~(\ref{b2mod2}).  It turns out that proceeding in the other direction has a complication that needs explanation;  Simply differentiating one of the potentials wrt $u$ does not give Eq.~(\ref{b2mod2}).  Instead an auxiliary potential is also required.  Before doing so it is convenient to define $h_-(x)=x-1-\sqrt{(x-1)^2+x/Q^2}$, $\beta_-=g_2 h_-/2k_B$, $u_+=2Q^2\langle b^2\rangle_0 h_+$, and $u_-=2Q^2\langle b^2\rangle_0 h_-$.  Both $u_+$ and $u_-$ are solutions to the quadratic equation formed when setting $du/dt=0$.  However, only $u_+$ makes physical sense as the stationary state.  The auxiliary potential $\Psi_-$ is defined as $\Psi_-=\Psi_{2,0}+k_B\beta_- u -\frac{g_{GT}}{8Q^2\langle b^2\rangle_0}u^2$.  Creating $\Psi_-$ poses no extra difficulties after $\Psi_2$ is created.  Towards obtaining $du/dt$ from potentials, noting that $\partial\Psi_-/\partial u=-\frac{g_2}{4Q^2\langle b^2\rangle_0}(u-u_-)$ which is positive definite for the range of physically allowable $u$.  Using both potentials gives the dynamics as
\begin{equation}
\frac{du}{dt}=-\frac{2L_{GT}^2}{c^2(\tau^*)^2}\left(\frac{\partial\Psi_-}{\partial u}\right)\left(\frac{\partial\Psi_2}{\partial u}\right)       ~\, . \label{dudtPsi}
\end{equation}

In contrast, an auxiliary potential is not required in the infinite $Q$ approach where the dynamics for $b$ comes directly from $\Psi_{2,\infty Q}\equiv \tau^*(1-z^{-2})\Phi_{2,\infty Q}$ and 
\begin{equation}
\frac{db}{dt}=\frac{L_{GT}}{2}\left(\frac{\partial\Psi_{2,\infty Q}}{\partial b}\right)       ~\, . \label{dbdtPsi}
\end{equation}

\subsection{Nonequilibrium Free Energy Functional}\label{sec:noneqfunc}

The above results are easily transferable to the treatment of order parameter fields, where $b(t)$ is replaced by $\phi(\bold{r},t)$, which varies spatially. For TI1, one uses the variable $\langle \phi(\bold{r},t)\rangle$. For example, for the first order free entropy potential, one uses $\Psi_1(\langle \phi(\bold{r},t)\rangle)$, which gets abbreviated for convenience to $\Psi_1(\phi(\bold{r},t))$. For second order, the variable to use is $\psi(\bold{r},t)\equiv\langle \phi^2(\bold{r},t)\rangle-\langle \phi^2(\bold{r})\rangle_0$, and the TI2 free entropy potential is $\Psi_2(\psi(\bold{r},t))$. 

The nonequilibrium free entropy is written as a functional of two functions $\phi$ and $\psi$ as $\Psi[\phi(\bold{r},t),\psi(\bold{r},t)]=\int{d^3 r \mathpzc{f}[\phi(\bold{r},t),\psi(\bold{r},t)]}$ with free entropy density $\mathpzc{f}(\phi(\bold{r},t),\psi(\bold{r},t))$. For the special case where $\gamma=0$ and $\langle b\rangle=0$, a prototype form of
\begin{equation}
\mathpzc{f}_p=\mathpzc{f}_0-k_B T\beta_2\psi(\bold{r},t)+\frac{Tg_{GT}}{8Q^2\langle b^2\rangle_0}\psi^2(\bold{r},t) \, \label{proto}
\end{equation} 
borrowed from Eq.~(\ref{psi2b}) will work.

For TI2 in the infinite $Q$ scheme, the $\phi$ field is used with $\Psi_{2,\infty Q}[\phi(\bold{r},t)]=\int{d^3 r \mathpzc{f}_{2,\infty Q}[\phi(\bold{r},t)]}$ 
%$\mathpzc{F}_{2,\infty Q}[\phi(\bold{r},t)]=U-TS+\int{d^3 r \mathpzc{f}_{2,\infty Q}[\phi(\bold{r},t)]}$ 
where 
%$\mathpzc{f}_{2,\infty Q}[\phi(\bold{r},t)]=-T\Psi_{2,\infty Q}[\phi(\bold{r},t)]$ and
\begin{equation}
\mathpzc{f}_{2,\infty Q}[\phi(\bold{r},t)]=L_{DR}(X_{DR}^2-X_c^2)\tau^* +g_2(z^2-1)\phi^2-\frac{1}{2}g_4 \phi^4 \,. \label{Psiphi}
\end{equation} 
Modifying Eq.~(\ref{dbdtPsi}) for a functional derivative gives the dynamics for the order parameter in this scheme is
%\begin{equation}
%\frac{\partial\phi}{\partial t}=\frac{(z^2-1)}{\tau_{GT}}\phi-\frac{g_4}{g_2\tau_{GT}}\phi^3 ~\, . \label{dphidtQ}
%\end{equation}
\begin{equation}
\frac{\partial\phi}{\partial t}=\frac{(z^2-1)}{\tau_{GT}}\phi-\frac{1}{4Q^2\langle \phi^2\rangle_0\tau_{GT}}\phi^3 ~\, . \label{dphidtQ}
\end{equation}

Equation~(\ref{dphidtQ}) resembles the SH equation.
The SH equation is an important tool for researchers of nonequilibrium systems and has, for example, been successfully utilized to model patterns formed in laminar flame fronts, certain types of Poiseuille flow, trapped ion modes in plasmas, and systems with Eckhaus instabilities such as Rayleigh-B\'enard convection~\cite{Kuramoto1976,SIVASHINSKY1977,MICHELSON1986,Manneville1988,Cross1993,Desai,Cross}.        The most important feature of the SH equation is a tunable parameter $r$ which can produce a negative relaxation rate, and instability, at $r=1$ in
\begin{equation}
\frac{\partial\phi}{\partial t}=(r-1)\phi -2\nabla^2 \phi -\nabla^4 \phi -\phi^3   ~\,. \,\,\,\,\,\,\,\,\, \text{(SH equation)}   \label{SH}
\end{equation}
In Eq.~(\ref{dphidtQ}) the first term on the right-hand side is the vital component in the SH equation, with $X_{DR}$ being the tunable parameter.   
A nonequilibrium PT occurs when $X_{DR}^2$ exceeds $X_c^2$, or equivalently, $r>1$.

The key differences between Eq.~(\ref{dphidtQ}) and the SH equation are the derivative terms, in particular a type of negative diffusion term, $-2\nabla^2 \phi$.  Dealing with such terms requires an analysis for more than one gate variable which is the topic of Sec.~\ref{sec:more}.  Before doing so, one can add some simple physics for the order parameter as either diffusion or as a surface tension (or interface energy) term to modify Eq.~(\ref{dphidtQ}) to
\begin{equation}
\frac{\partial\phi}{\partial t}=\frac{(z^2-1)}{\tau_{GT}}\phi+D\nabla^2\phi-\frac{1}{4Q^2\langle \phi^2\rangle_0\tau_{GT}}\phi^3  ~\, . \label{LangevinA}
\end{equation}
The form of Eq.~(\ref{LangevinA}) is identical to the Langevin Model A equation (without noise), under the Hohenberg-Halperin classification scheme used to describe the (nonconserved) order parameter critical dynamics, including quench dynamics, near an equilibrium PT~\cite{Desai}.  This comparison works right down to the term $z^2-1$ playing the role of $T-T_c$, and changing sign at the critical point.  I point out that the physics of this comparison is unclear, as far whether or not TI2 would play any role near an equilibrium PT.  At the very least, the TI2 approach presented here could play a useful role in modeling quench dynamics near equilibrium PTs.   
Using $\psi$ instead of $\phi$ will account for fluctuations in the order parameter field.  The following adaptation of Eq.~(\ref{b2mod2}) (plus adding in diffusion): 
%\begin{equation}
%\frac{\partial\psi}{\partial t}=4\alpha L_{GT}\tau^*\left(X_{DR}^2-X_c^2\right)\psi+4\alpha L_{GT}\tau^* X_{DR}^2\langle b^2\rangle_0 +D_u\nabla^2\psi  -4c^2 \psi^2 ~\,,   \label{dpsidt}
%\end{equation}
\begin{equation}
\frac{\partial\psi}{\partial t}=2\frac{\left(z^2-1\right)}{\tau_{GT}}\psi+2\frac{z^2}{\tau_{GT}}\langle b^2\rangle_0 +D_u\nabla^2\psi  -2c^2 \psi^2 ~\,,   \label{dpsidt}
\end{equation}
will provide just as good a model for describing quenching as the Langevin Model A equation.  Near the critical point $\psi$ becomes large and thus describes well the fluctuations that would be created by adding in a noise term to the Langevin Model A.

\section{Case of more than one gate variable} \label{sec:more}

Many of the results presented in Sec.~\ref{sec:gentheory}  are easily transferable to the case of more than one gate variable. 
% I will discuss in detail the case for TI2 with a dispersion relation for the excitation gap, and then the specific case of TI1 on momentum variables.
%\subsection{Second Order TI}
For the case where there are $N$ identical copies of the gate system considered in Sec.~\ref{sec:gentheory}, each gate would have the same values of $\alpha$, $X_c$, and $Q$.
Having more gate systems has the obvious advantage of boosting the entropy budget available for creating patterns.  The expression in Eq.~(\ref{SGTquad}) is multiplied by $N$; In particular at critical, $(z=1)$, $\Delta S_{GT,ss}=-2NQk_B$.  If one is interested in patterns being used to store information, then this entropy budget is sufficient to store up to $2NQ/\ln2$ bits of information.  How the storing of this information occurs is not a trivial matter and will depend on the pattern quality which involves characteristics such as the long range order.
%If there is no $q$ dependence for $\alpha_q$, $\tau_q^*$, $L_q$, and $c_q$ then all gate states are effectively degenerate and there will be just one transition.  
If $q$ physically represents a wavevector then all length scales will be excited with equal weight.  If all modes are excited equally then long range order is not expected and clear patterns are unlikely to form.  This analysis is carried out below in Sec.~\ref{fluctmany}.
% resulting in scaling behavior.  These spatial excitations in general represent Goldstone modes, each breaking symmetry at the particular wavevector above critical. 
% There is nothing in this analysis to fix the phase of these modes so one must assume that there would be no phase coherence between any two modes.  Also the phase angle of any mode would likely fluctuate in time.  Above critical the system would appear to be noisy and/or turbulent in nature.  If the physical system lacks translational symmetry to begin with then the phase fluctuations could be reduced.  For example, if the container for the system has even a slight inherent undulation at wavevector $q$ then the phase angle of the $q$ mode could lock-in and stabilize.
Thus, pattern formation requires $\alpha$ to depend on $q$.

%  If there is a pattern to be observed from the bifurcation, then having $N$ systems bifurcating should boost the pattern by having the signal to noise improve by a factor of $N^{1/2}$.
%The lineshape function mu can be exploited as well.  For $N=1$, with $z=1$, $u=2Q+1$ times b02.  If this happens to be the condition for discerning some type of pattern, then with N identical gate variables then the same condition can be met at a lower z, i.e., at z  For large N this can be well below the excitation gap.  This could be very helpful in systems where for various reasons $X_c$ cannot be reached.  By having large redundancy a pattern can be formed well below the true threshold.  This may be the case in the brain if each neuron can be identified as a gate variable.  It seems reasonable that as organisms developed with primitive brains, the chemical threshold for Xc may have been too large to attain, so these organisms compensated by having more neurons.
%can get below threshold patterns if the dynamics are slow enough.  If the instruments used to observe the system are faster than the gate dynamics then the pattern seen and will not be seen as mere noise.  

When variables $\alpha$, $g_{GT}$, and $c$ depend on $q$, the bifurcation point $X_c$ depends on $q$ with a relationship I call the dispersion relation.  Equation~(\ref{Xcrit}) becomes
\begin{equation}
X_{c,q}= \sqrt{\frac{g_{q}}{2\alpha_q \tau^*}}        ~\, . \label{Xcritq}
\end{equation}
Since the parameters $g_q$ and $\alpha_q$ must always remain positive it is natural to define the \textit{excitation gap}, $X_{eg}$, as the minimum value of $X_{c,q}$.  
%This gap is important because when $X_{DR}$ is raised upwards from zero, the first critical point to be reached is at the excitation gap with mode $q_{eg}$.
As one ramps up $X_{DR}$ from zero there will be an interval, this excitation gap, over which little happens before $X_{DR}$ reaches the first of a set of discrete levels at $X_{c,q}$, each one creating a transition at a given mode $q$ that makes $b_q$ nonzero.  This is analogous to quantum transitions in molecular systems when the excitation energy matches or exceeds the energy gap between highest occupied and lowest unoccupied molecular orbitals, or in solid state insulators where energy levels coalesce to form bands, and the excitation energy meets or exceeds the band gap energy.  This point of view is in the infinite $Q$ limit.  For finite $Q$, there will be enhanced fluctuations as $X_{DR}$ approaches the gap.  In either case, when the excitation gap is surmounted one might expect some type of pattern to form in the system with a Fourier transform peaked around $q_{eg}$.  This patterned system could also be referred to as a self-assembled or self-organized structure. 

The potentials $\Phi_1$, $\Phi_{2,Q\infty}$, and $\Phi_2$ are straightforward to generate from Eqs.~(\ref{Phi1}) , (\ref{Phi2}), and (\ref{Psi2mod})
\begin{equation}
\Phi_1=L_{DR}X_{DR}^2+\sum_{q}\left[X_{DR}^2\gamma_q b_q -\lambda_{1,q}\left(-g_{q}\gamma_q  L_{q}\tau^*X_{DR}^2b_q +L_{q}g_{q}^2 b_q^2 \right)\right]   ~\, , \label{Phi1sum}
\end{equation}
\begin{equation}
\Phi_{2,\infty Q}=L_{DR}X_{DR}^2+\sum_{q}\left[X_{DR}^2\alpha_q b_q^2 -\lambda_{Q\infty,q} \left(-2\alpha_q \left(X_{DR}^2 -X_{c,q}^2\right) L_{q}g_{q}\tau^* b_q^2 +c_q^2 g_{q} b_q^4 \right)\right]   ~\, , \label{Phi2sum}
\end{equation}
\begin{equation}
\Phi_2(u)=L_{DR}X_{DR}^2+\sum_{q}\left[\frac{g_{q}\mu}{2\tau^*}\left(z_q^2\langle b_q^2\rangle_0 +u_q h_+(z_q^2)-\frac{u_q^2}{4Q_q^2\langle b_q^2\rangle_0}\right)\right] ~\, , \label{Phi2usum}
\end{equation}
where $\lambda_{1,q}=\tau_q/\tau^*$, $\lambda_{Q\infty,q}=\tau_q X_{DR}^2/2\tau^*(X_{DR}^2-X_{c,q}^2)$, and $z_q^2=X_{DR}^2/X_{c,q}^2$.
When constructing the free entropy potentials it is best to first drop the $L_{DR}X_{DR}^2$ terms in the $\Phi$ potentials and then divide each term by the required factors including the Lagrange multipliers.  Dropping the $L_{DR}X_{DR}^2$ terms makes no difference in equilibrium and also when differentiating by any variables $b_q$ and $u_q$.  This construction yields:
\begin{equation}
\Psi_{1}=\sum_{q}\left[k_B\beta_{1,q}b_q -\frac{1}{2}g_{q}b_q^2\right]    ~\, , \label{psi1bsum}
\end{equation}
\begin{equation}
\Psi_{2,\infty Q}=\sum_{q}\left[(z_q^2-1)g_qb_q^2 -\frac{1}{2} g_{4,q} b_q^4 \right]   ~\, , \label{Psi2Qsum}
\end{equation}
%check minus sign on g4 -yes checked
and
\begin{equation}
\Psi_2=\frac{1}{2}\sum_{q}\left[g_qz_q^2\langle b_q^2\rangle_0 +g_qu_q h_+(z_q^2)-g_{4,q} u_q^2\right] ~\, . \label{Psi2usum}
\end{equation}
As discussed below, these $q$ sums may or may not include $q=0$, depending on the physical details of the DR. 
Replacing $h_+$ with $h_-$ in Eq.~(\ref{Psi2usum}) gives the auxiliary potential $\Psi_-$ from which one may derive the dynamics for each $u_q$, following Eq.~(\ref{dudtPsi}):
\begin{equation}
\frac{du_q}{dt}=-\frac{2}{(cg_q\tau_q\tau^*)^2}\left(\frac{\partial\Psi_-}{\partial u_q}\right)\left(\frac{\partial\Psi_2}{\partial u_q}\right)       ~\, . \label{duqdtPsi}
\end{equation}
Equation~(\ref{duqdtPsi}) accounts for fluctuations of many modes $q$ near a PT, and represents the most complete description to date of nonequilibrium dynamics in a system that breaks symmetry at a nonequilibrium PT.

 From $\Psi_1$ and $\Psi_2$, one constructs $\Psi=\Psi_1+\Psi_2$, $\mathpzc{S}_{NE}$ and $\mathpzc{F}$, which obey general extremum principles by stationary states (one for each $q$).

\subsection{Case where $\alpha_q=\alpha_0+\alpha_2 q^2 +\alpha_4 q^4$}\label{sec:alpha0}

I now show that the SH equation can be derived from first principles via the infinite $Q$ approach inside of TI2, if the induction parameter $\alpha_q$ has the following small $q$ expansion up to fourth order:
\begin{equation}
\alpha_q=\alpha_0+\alpha_2 q^2 +\alpha_4 q^4  ~\,.   \label{alphaq}
\end{equation}
The starting point is the linearized (unconstrained) Eq.~(\ref{GTdyn2}) modified for each mode $q$:
\begin{equation}
\dot{b}_q=\frac{z_0^2-1}{\tau_{GT}} b_q +D_{TI} q^2 b_q +d_4 q^4 b_q~  \,,       \label{SHa}
\end{equation}
where $z_0^2=2\tau^* X_{DR}^2\alpha_0/g_{GT}$, $D_{TI}=2L_{GT}\tau^* X_{DR}^2\alpha_2$, and $d_4=2L_{GT}\tau^* X_{DR}^2\alpha_4$.  As the $b_q$ are Fourier components of $\phi(\bold{r})$ then the Fourier transform of Eq.~(\ref{SHa}) gives the desired result after diffusion with coefficient $D$ is taken into account, and the cubic cut-off term is returned:
\begin{equation}
\frac{\partial\phi}{\partial t}=\frac{z_0^2-1}{\tau_{GT}}\phi +(D-D_{TI}) \nabla^2 \phi +d_4 \nabla^4 \phi -\frac{1}{4Q^2\langle \phi^2\rangle_0\tau_{GT}}\phi^3~  \,.       \label{SHc}
\end{equation}
With negative $d_4$, Eq.~(\ref{SHc}) is the SH equation, Eq.~(\ref{SH}).
The induction parameter $D_{TI}$ acts like a negative diffusion coefficient.  This means that instabilities can be created by making $X_{DR}$ large enough and this can lead to either a negative effective relaxation rate, a negative effective diffusion coefficient, or both.  This has been shown to produce patterns in simulations~\cite{Cross}.  When returning to Fourier components, the condition for instability is
\begin{equation}
\frac{z_0^2-1}{\tau_{GT}}-(D-D_{TI}) q^2  +d_4 q^4 >0~  \,.       \label{SHcond}
\end{equation}
Solving for $X_{DR}$ at the instability threshold gives the nonequilibrium dispersion relation:
\begin{equation}
X_{DR}(q)=X_{c0}\sqrt{\frac{1+\tau_{GT}Dq^2}{\alpha_q/\alpha_0}}~  \,.       \label{DispSH}
\end{equation}
The dynamics for the SH equation can be obtained from a free entropy functional as 
%$\mathpzc{\psi}_{SH}=-\mathpzc{f}/T$   $\mathpzc{\phi}$
\begin{equation}
\frac{\partial\phi}{\partial t}=\frac{1}{2}\frac{\delta}{\delta\phi}\int{d^3 r\left[ \frac{z_0^2-1}{\tau_{GT}}\phi^2 -(D-D_{TI}) (\nabla \phi)^2 +d_4 (\nabla^2 \phi)^2 -\frac{1}{4Q^2\langle \phi^2\rangle_0\tau_{GT}}\phi^4 \right]}~ \,.  \label{SHfunct}
\end{equation}
%now SH eq with derivs $\alpha_q=\alpha_0+\alpha_2 q^2$

For the case of dissipative systems, any phenomenon successfully modeled with the SH equation is also successfully modeled by TI2.  When $D-D_{TI}<0$, the SH model on its own violates the SLT, but under TI2 the solution to the entropic coupling problem applies and the SLT is never violated. 

When dealing with fluctuations, Eq.~(\ref{b2mod2}) may be modified to:
\begin{equation}
\frac{1}{2}\frac{du_q}{dt}=\left(\frac{z_0^2}{\tau_{GT}}+D_{TI}q^2+d_4q^4\right)\left(u_q+\langle b^2\rangle_0\right) -\frac{u_q}{\tau_{GT}}-Dq^2u -c^2u_q^2 ~\,.   \label{uqdotalpha}
\end{equation}
A real space version can be produced as:
\begin{equation}
\frac{1}{2}\frac{\partial\psi}{\partial t}=\left(\frac{z_0^2}{\tau_{GT}}\psi+(D-D_{TI})\nabla^2\psi+d_4\nabla^4\psi\right)\left(\psi+\langle b^2\rangle_0\delta(\bold{r})\right) -c^2\psi^2 ~\,.   \label{psidotalpha}
\end{equation}
Equations~(\ref{uqdotalpha}) and (\ref{psidotalpha}) would describe the finite-Q broadened resonant transition as $X_{DR}$ is varied above the excitation gap.  Near the gap fluctuations are large and these equations should be useful in modeling any nonequilibrium phenomena with large fluctuations.  Such systems have been modeled using the Edwards-Wilkinson (EW), Burgers, complex Ginzberg-Landau, and Kardar-Parisi-Zhang (KPZ) equations, all of which have noise terms added in artificially~\cite{Desai,Cross,KPZ1986}.  I believe that the essential physics contained in all of these models can be arrived at from the basic TI analysis up to second order, leading up to Eqs.~(\ref{uqdotalpha}) and (\ref{psidotalpha}).  In Eqs.~(\ref{uqdotalpha}) and (\ref{psidotalpha}) noise is incorporated more naturally, with noise levels varying sensitively with $Q$ and $X_{DR}$.

%The KS equation bears a strong resemblance to other models used widely in the study of nonequilibrium systems, i.e., the Edwards-Wilkinson (EW), Burgers, complex Ginzberg-Landau, and Kardar-Parisi-Zhang (KPZ) equations.
%It would be interesting to see if the essential physics contained in all of these models can be arrived at from the basic TI analysis up to second order.
The EW and KPZ equations do not have a negative diffusion coefficient or a negative relaxation time so are not models corresponding to behaviour above a nonequilibrium critical point.
They also do not have higher order quenching terms.  They do, however, possess an external stochastic noise term.  The EW equation is simply the diffusion equation with an added noise term.
Since fluctuations are large near the critical point, $X_{eg}$, the EW equation and Eqs.~(\ref{uqdotalpha}), (\ref{psidotalpha}) are well-suited for describing nonequilibrium systems just below, and right up to, the critical point.
%Understanding that Eq.~(\ref{bqdot}) is a good approximation only well above critical, and only a rough approximation near critical, it is important near and below critical to use dynamical equations for $u_q=b_q^2-\langle b_q^2\rangle_0$, with dynamics similar to Eq.~(\ref{b2mod2}). The proper equation is:
%\begin{eqnarray}
%\dot{u}_q&=&2\left(\frac{X_{DR}^2}{X_{c}^2(q)}-1\right) \frac{u_q}{\tau_q}  +2\left(\frac{X_{DR}^2}{X_{c}^2(q)}\right) \frac{\langle b_q^2\rangle_0}{\tau_q} -4c_q^2 u_q^2 \\ \nonumber
 %&+&2Dq^2\left(\frac{X_{DR}^2}{X_{c}^2(q)}-1\right)u_q +2Dq^2\left(\frac{X_{DR}^2}{X_{c}^2(q)}\right)\langle b_q^2\rangle_0 -4d_q^2 q^2 u_q^2    ~\,.   \label{uqdot}
%\end{eqnarray}
  %Equation~(\ref{uqdot}) describes the fluctuations in the modes $b_q$ without having to artificially add in an external noise term.  While the amount of noise added into the EW equation is arbitrary, the amount of fluctuations in $b_q$ depends on $X_{DR}$, with a narrow, peaked function depending on the lineshape function $\mu$, as indicated in Eq.~(\ref{uquad}).  Equation~(\ref{uqdot}) is a good description of the thermodynamics of bifurcation in a real system containing many wavevector modes.
%The mean position of the interface is described by the DR variable $b_0$ and Eq.~(\ref{b0dotKS}).
%The lineshape function mu must be invoked and the dynamics must focus on b2
A nonequilibrium PT may be an explanation for the kinetic roughening transition sometimes observed in film growth and modeled by the EW and KPZ equations~\cite{Mukamel1998,Pinnington1997,Desai}.  A high influx of adsorbates, and thus a large $X_{DR}$, would be required to achieve this type of nonequilibrium PT.  Above the bifurcation point, the EW equation is not suitable, whereas Eqs.~(\ref{uqdotalpha}) and (\ref{psidotalpha}) are well-suited for this task, as the adjustable reactant flux in film growth flux is related to $X_{DR}$.

%\subsection{case where $\alpha_q=q^2\eta_q$}\label{sec:eta}
\subsection{Case of Conserved Order Parameter}\label{sec:consord}
%\subsection{$q=0$ mode as DR}\label{sec:q0DR}

In systems with conserved order parameters, direct relaxation has little influence and Eq.~(\ref{SHc}) becomes:
%\begin{equation}
%\frac{\partial\phi}{\partial t}= +(D-D_2) \nabla^2 \phi +d_4 \nabla^4 \phi -\frac{1}{4Q^2\langle \phi^2\rangle_0\tau_{GT}}\phi^3~  \,.       \label{LangB}
%\end{equation}
\begin{equation}
\frac{\partial\phi}{\partial t}= \nabla^2\left[(D-D_{TI})  \phi +d_4 \nabla^2 \phi -\frac{1}{4Q^2\langle \phi^2\rangle_0\tau_{GT}}\phi^3\right]~  \,.       \label{LangB}
\end{equation}
The extra $\nabla^2$ operator arises from the relation $\frac{\partial\phi}{\partial t}=-\del\cdot\bold{j}$ while the current density is given by $\bold{j}=-M\del\mu$.  The chemical potential $\mu$ can be expressed as $\mu(\bold{r},t)=\delta\mathpzc{F}/\delta\phi$ where
\begin{equation}
\mathpzc{F}=\frac{1}{2M}\int{d^3 r\left[ (D-D_{TI})\phi^2 -d_4 (\nabla \phi)^2 -\frac{1}{4Q^2\langle \phi^2\rangle_0\tau_{GT}}\phi^4 \right]}~  \,.       \label{LangBfunct}
\end{equation}
Mathematically, Eqs.~(\ref{LangB}) and (\ref{LangBfunct}) coincide with the Langevin B Model used to model the interfacial structure created after quenches through equilibrium PTs.  This model is built from the Landau-Ginzberg-Wilson free energy functional, $\mathpzc{F}_{LGW}$.  In this case  the first coefficient $(D-D_{TI})$  changes sign at the transition temperature and the physics behind this is well understood without invoking TI.  Even though it is tempting to associate $X_{DR}$ with $(T_c-T)/T_c^2$ in the specific quenching case discussed here, the physical reasoning for any induction is unclear.  As things stand, Eqs.~(\ref{LangB}) and (\ref{LangBfunct}) should be only thought of as an effective model for PTs with conserved order parameters.  From the modeling point of view, Eqs.~(\ref{uqdotalpha}) and (\ref{psidotalpha}) may provide better results than Eq.~(\ref{LangBfunct}) since they better account for fluctuations. 

%\subsection{Variational Principles}

Returning to fundamental thermodynamics, it is not surprising that the Landau-Ginzberg-Wilson free energy functional, $\mathpzc{F}_{LGW}$ has been used to model nonequilibrium phenomena, given the success of using this functional towards understanding second order equilibrium PTs~\cite{Desai}.  However the physical significance of minimizing this functional has never been established since $\mathpzc{F}_{LGW}$ is the Helmholtz free energy and can only be minimized in thermodynamic equilibrium.  The reality is that $\mathpzc{F}_{LGW}$ has been used because there were no other options available.  Thermodynamic functionals that have extrema at nonequilibrium PTs must be different, and I have shown here how these functionals are constructed, with $\mathpzc{F}$ being the best substitute for $\mathpzc{F}_{LGW}$. 

\subsection{Case where the DR is the $q=0$ mode}

At this point I will emphasize important systems that approach equilibrium through diffusion, i.e., where the relaxation rate $1/\tau_{GT}$ is effectively $Dq^2$.  In particular I focus on systems where the complete set of $q$ modes acts as both DR and gate.  The $q=0$ mode is the DR and all other wavevectors are gate states. To make a clear demarcation between DR and gate, the induction must disappear as $q$ approaches zero.  For this reason I consider only systems with $\alpha_0=0$ and I consider specifically the restricted case where $\alpha=\alpha_2 q^2$.  These results will be applicable to two important physical examples discussed below in Sec.~\ref{sec:front} and Sec.~\ref{sec:turing}.  
%In these examples I will also point out what distinguishes the gate and SI.  Though the modes $q\neq 0$ are the SI, the SI  For now, the terms gate and SI are 

The DR dynamics is given by
\begin{equation}
\dot{b}_0=L_{DR}X_{DR}+\frac{1}{2}\alpha_2X_{DR} \sum_{q} q^2 b_q^2    ~\,.   \label{b0dotgen}
\end{equation}
%This is an example of the nonequilibrium Le Chatelier's principle.
For the variables describing the gate, Eq.~(\ref{GTdyn2}) becomes, 
%up to linear order:
%\begin{equation}
%\dot{b}_q=2\alpha_q L_{GT} \tau^* X_{DR}^2 b_q  - Dq^2 b_q -c_q b_q^3  ~\,.   \label{bqdotFront}
%\end{equation}
%\begin{equation}
%\dot{b}_q=2\alpha_q L_{GT} \tau^* X_{DR}^2 b_q  - Dq^2 b_q   ~\,.   \label{bqdotFront}
%\end{equation}
%\begin{equation}
%\dot{b}_q=2\frac{\alpha_2}{g_q} Dq^4 \tau^* X_{DR}^2 b_q -Dq^2 b_q-c_q^2 b_q^3   ~\,.   \label{bqdotD}
%\end{equation}
\begin{equation}
\dot{b}_q=2\frac{\alpha_2}{g_2} Dq^4 \tau^* X_{DR}^2 b_q -Dq^2 b_q-\frac{g_4}{g_2}Dq^2 b_q^3   ~\,.   \label{bqdotD}
\end{equation}
Equations~(\ref{b0dotgen}) and (\ref{bqdotD}) are extensively coupled in general since $X_{DR}=-g_{DR}b_0$.  However when the DR is very large and slow, $X_{DR}$ is considered to be essentially constant in Eq.~(\ref{bqdotD}).
%The result is a variable effective diffusion coefficient $D-D_{TI}$ which can become negative with large enough $X_{DR}$ ,i.e., at $X_c(q)=\sqrt{g_q/L_{DR}\tau^*q^2}$. 
There is an issue with Eqs.~(\ref{b0dotgen}) and (\ref{bqdotD}) having a discontinuity at $q=0$.  From Eq.~(\ref{bqdotD}) $\dot{b}_q$ is very small for small $q$, while $\dot{b}_0$ may in fact be quite large.  The best way to deal with this is to use the well-known procedure of redefining the real-space variable $\phi$ as following the ''front'' as done in the formulation of the KPZ equation~\cite{Desai}.  To linear order this is done by adding the spatially uniform term $L_{DR}X_{DR}t$ to $\phi$.  This works well if $L_{DR}X_{DR}$ varies slowly in time and this is indeed the case when studying short-term front dynamics. One can refine this by adding the (also spatially uniform) term $\frac{1}{2}\alpha_2X_{DR} \sum_{q} q^2 b_q^2$ as well, to follow the growth more accurately.  Explicitly then the new front-following field is $\phi'=\phi-(L_{DR}X_{DR}+\frac{1}{2}\alpha_2X_{DR} \sum_{q} q^2 b_q^2)t$.  For this front-following procedure, a spatial average is implied, so the dynamics for the $b_q$ modes in Eq.~(\ref{bqdotD}) is unaffected, while for the primed variable Eq.~(\ref{b0dotgen}) for the $q=0$ mode becomes a trivial identity.  %Also, higher order induction terms would need to be added to Eq.~(\ref{b0dotgen}).  In any case, 
This procedure will work just as well on systems without a physical front/interface, for example, when $b_0$ represents the extent of reaction during a chemical reaction.
When taking the Fourier transform of $b_q$ (see below), Eq.~(\ref{bqdotD}) does indeed have $\dot{b}_q$ continuously going to zero as $q$ approaches zero.  For further analysis below the prime on $\phi'$ is dropped for convenience and $\phi$ is to be understood as tracking the front.

The appropriate potentials at this level of analysis are the infinite $Q$ potentials $\Phi_{2,\infty Q}$ and $\Psi_{2,\infty Q}$, each functions of the set $b_q$.  
For $\Phi_{2,\infty Q}(b)$, Eq.~(\ref{Phi2sum}) becomes:
% Note replace tauGT with 1/Dq2
\begin{equation}
\Phi_{2,\infty Q}=L_{DR}X_{DR}^2+\sum_{q\neq 0}{\left[\alpha_2X_{DR}^2 q^2b_q^2 -\lambda_q \left(-2\alpha_2 \left(X_{DR}^2 -X_{c,q}^2\right)Dq^4\tau^* b_q^2 +g_4 Dq^2 b_q^4 \right)\right]}   ~\, , \label{Phi2q}
\end{equation} 
where $X_{c,q}=\sqrt{g_2/2\alpha_2\tau^*q^2}$, $\lambda_q=0$ when $X_{DR}<X_{c,q}$, and $\lambda_{q}= (1/2Dq^2\tau^*)(X_{DR}^2/(X_{DR}^2-X_{c,q}^2)$ when $X_{DR}\geq X_{c,q}$.
Below the threshold for each $q$ there is no induction in the infinite $Q$ limit.  Defining $q_c\equiv\sqrt{g_2/2\alpha_2\tau^*X_{DR}^2}$, the sum $\sum_{q\neq 0}$ in Eq.~(\ref{Phi2q}) may be replaced by $\sum_{q>q_c}$.  The corresponding free entropy, from Eq.~(\ref{Psi2Qsum}) is
%Towards establishing $\Psi_2$, note that the first term in Eq.~(\ref{Phi2q}) disappears in equilibrium and has no bearing on any variational procedures.  I drop this term and multiply each term in the $q$ sum by $(\tau^*)(X_{DR}^2-X_{c,q}^2)/X_{DR}^2$ to give
%\begin{equation}
%\Psi_{2,\infty Q}=\sum_{q\neq 0}{\left[\alpha_2(X_{DR}^2-X_{c,q}^2)\tau^* q^2b_q^2 -\left(-\alpha_2 \left(X_{DR}^2 -X_{c,q}^2\right)q^2\tau^* b_q^2 +\frac{1}{2}g_4 b_q^4 \right)\right]}   ~\, , \label{Psi2q}
%\end{equation}
%\begin{equation}
%\Psi_{2,\infty Q}=\sum_{q\neq 0}{\left[2\alpha_2(X_{DR}^2-X_{c,q}^2)\tau^* q^2b_q^2 -\frac{1}{2}g_4 b_q^4 \right]}   ~\, , \label{Psi2q}
%\end{equation}
\begin{equation}
\Psi_{2,\infty Q}=\sum_{q>q_c}{\left[2\alpha_2X_{DR}^2\tau^* q^2b_q^2-g_2b_q^2 -\frac{1}{2}g_4 b_q^4 \right]}   ~\, . \label{Psi2q}
\end{equation}

If the linearized version of Eq.~(\ref{bqdotD}) is Fourier transformed and an appropriate large amplitude cutoff is re-introduced then the following PDE is:
\begin{equation}
\frac{\partial\phi}{\partial t}=2\frac{\alpha_2}{g_2}D\tau^* X_{DR}^2\nabla^4\phi+D\nabla^2\phi+\frac{g_4}{g_2}D\nabla^2\phi^3   ~\,.   \label{phidotD}
\end{equation}
A functional defined as:
\begin{equation}
\Psi_{2,\infty Q}[\phi(\bold{r},t)]\equiv-\int{d^d \bold{r}\left[2\alpha_2X_{DR}^2\tau^*(\del\phi)^2+g_2\phi^2+\frac{1}{2}g_4 \phi^4 \right]}   ~\, , \label{Psi2qfunct}
\end{equation}
is appropriate for determining nonequilibrium dynamics by taking one functional derivative with $X_{DR}$ held constant:
\begin{equation}
\frac{\partial\phi}{\partial t}=\frac{2g_2}{D}\nabla^2\left(\frac{\delta\Psi_{2,\infty Q}}{\delta\phi}\right)   ~\, . \label{FunctDeriv}
\end{equation}
A time-independent function $\phi(\bold{r})$ that maximizes $\Psi_{2,\infty Q}$ will also be a stationary state.
%\begin{equation}
%\Phi_{2,\infty Q}(b)=\sum_{q\neq 0}{X_{DR}^2\alpha b_q^2} -\sum_{q\neq 0}{\lambda_q \left[-2\alpha \left(X_{DR}^2 -X_{c,q}^2\right) L_{GT}g_{GT}\tau^* b^2 +c_q^2 g_{GT} b^4 \right]}   ~\, , \label{Phi2Turing}
%\end{equation} 
%
%\begin{equation}
%\Phi_{2,\infty Q}(b)=\sum_{q\neq 0}{\left[\frac{g_q}{2\tau^*}\frac{A^2}{A_c^2}b_q^2 +\lambda_q g_q\left(\frac{A^2}{A_c^2}-1\right)Dq^2 b_q^2 -\lambda_q g_qc_q^2 b^4 \right]}   ~\, , \label{Phi2Turing}
%\end{equation}
%where $\lambda_q$ above critical is     
%\begin{equation}
%\lambda_{q}= \frac{\tau_{GT}}{2\tau^*}\frac{X_{DR}^2}{X_{DR}^2-X_{c,q}^2}   ~\, . \,\,\,\,\, \text{(above critical)} \label{lambdaq}
%\end{equation}
%\begin{equation}
%\lambda_{q}= \frac{1}{2\tau^* Dq^2}\frac{A^2}{A^2-A_{c,q}^2}   ~\, . \,\,\,\,\, \text{(above critical)} \label{lambdaq}
%\end{equation}
When dealing with fluctuations, one adapts Eq.~(\ref{b2mod2}) to become
\begin{equation}
\frac{du_q}{dt}=4\frac{\alpha_2}{g_2}\tau^*\left(X_{DR}^2-X_c^2\right)Dq^4u_q+4\frac{\alpha_2}{g_2}\tau^* X_{DR}^2Dq^2\langle b^2\rangle_0  -2\frac{g_4}{g_2}Dq^2 u_q^2 ~\,.   \label{uqdyn}
\end{equation}
In terms of the real space fluctuation function $\psi(\bold{r},t)$:
\begin{equation}
\frac{\partial\psi}{\partial t}=2D\nabla^2\left[2\frac{\alpha_2}{g_2}\tau^*X_{DR}^2\nabla^2\psi+\psi+2\frac{\alpha_2}{g_2}\tau^* X_{DR}^2\langle b^2\rangle_0\nabla^2\delta(\bold{r}) +\frac{g_4}{g_2}\psi^2\right] ~\,.   \label{psidyn}
\end{equation}

%In terms of the spatially averaged order parameter field:
%\begin{equation}
%\frac{d\widetilde{\phi}}{dt}=L_{DR}X_{DR}+\frac{1}{2}\alpha_2X_{DR}\widetilde{(\del\phi)^2}    ~\,.   \label{b0dotreal2}
%\end{equation}

In this section I have established the first principles theory required to treat two important physical examples, discussed next.  In these systems relaxation occurs by diffusion with rate $Dq^2$, the complete set of $q$ modes acts as both DR and gate, and there is a sound physical basis for having nonzero $\alpha_2$.

\subsection{Front Propagation}\label{sec:front}

Phase separation is common in nonequilibrium systems and the interface/front between the two phases will be expected to move as one phase overtakes the other.  These fronts can form after a quench occurs through an equilibrium PT.  The velocity $v$ of the front, when spatially averaged, will follow the typical dissipative relation:
\begin{equation}
v= M_{DR}X_{DR} ~  \,,       \label{v}
\end{equation}
where $X_{DR}=-\Delta\mu/T$ and $\Delta\mu$ is the chemical potential difference between the two phases in question. 
%the Onsager coefficient $M_{DR}$ may be derived from first principles using approaches such as the Allen-Cahn equation.
If the mean front velocity is in the $z$ direction, and $\chi(y,z)$ is the local position of the front then purely geometric effects arising from the eikonal factor $\sqrt{1+(\del\chi)^2}$ play an important role in the dynamics~\cite{Desai,Kuramoto1976}.  Incorporating this effect into the rate of front motion results in the following expression for the Onsager coefficient for each Fourier mode $q$:
%\begin{equation}
%M_{DR}=L_{DR}\left(1+\frac{1}{2}(\del\chi)^2\right) ~  \,,       \label{eikonal}
%\end{equation}
\begin{equation}
M_{DR,q}=L_{DR}\left(1+\frac{1}{2}q^2b_q^2\right) ~  \,,       \label{eikonal}
\end{equation}
where the small amplitude approximation was made, and the $b_q$ are Fourier coefficients of $\chi$.  The eikonal factor means that curvature in the interface will enhance the spatially averaged dynamics.  This enhancement attenuates at long wavelengths.  From the definition of $\alpha_q$ the eikonal factor directly results in TI2 with
\begin{equation}
\alpha_{q}=\frac{1}{2}L_{DR}q^2 ~  \,.       \label{eikalpha}
\end{equation}
Thus, the following structure emerges: 1) the $q=0$ mode is the DR variable, 2) all modes with $q\neq 0$ are gate modes, and 3) the DR and gates are distinct in this scheme because the TI2 effect goes as $q^2$ and gate states near $q=0$ are affected very little.  In this system both the DR and gate are built into the same continuous variable describing the front.
The DR variable is the overall (spatially averaged) position of the front, which is being driven by the chemical potential difference between the two competing phases.  Equation~(\ref{b0dotgen}) represents not only the overall motion of the front but also the considerable rate at which entropy is being created as one phase turns into another.  It is precisely this rate of entropy increase that prevents the SLT from being violated if any patterns were to form in the shape of the front.
The gate variables in this case describe the shape of the propagating front.  As the front propagates, the front will be exchanging particles with both phases (the particle reservoir) and possibly heat, with again the phases on both sides of the front.  Thus, it is the shape of the front and all of this exchanging that is actually the gate, which winds up being almost all of the system, in this example.
Equation~(\ref{eikalpha}) represents the entropic coupling between the DR and the gate modes, $q\neq 0$.

All of Eqs.~(\ref{b0dotgen}) through to (\ref{psidyn}) apply to this case with $\alpha_2=\frac{1}{2}L_{DR}$.
In particular the DR dynamics is given by Eq.~(\ref{b0dotgen}) while in real space the dynamics of 
%\begin{equation}
%\dot{b}_0=L_{DR}X_{DR}+\frac{1}{2}L_{DR}X_{DR} \sum_{q} q^2 b_q^2    ~\,.   \label{b0dotKS}
%\end{equation}
the spatially averaged front position is:
\begin{equation}
\frac{d\widetilde{\chi}}{dt}=L_{DR}X_{DR}\left(1+\frac{1}{2}\widetilde{(\del\chi)^2}\right)    ~\,.   \label{b0dotreal}
\end{equation}
As pointed out in the literature, the front velocity is always greater when undulations in the front, or interface, are taken into account~\cite{Desai}.  This is an example of the nonequilibrium Le Chatelier's principle.
 
When treating the dynamics for the gate states, the relaxation of undulations in the front shape occurs by diffusion only and the relaxation rate is $Dq^2$.  When Eq.~(\ref{bqdotD}) is linearized and expressed as $\dot{b}_q=(D_{TI}-D)b_q$ where $D_{TI}=L_{DR}Dq^2\tau^*X_{DR}^2/g_2$. The result is a variable effective diffusion coefficient $D-D_{TI}$ which can become negative with large enough $X_{DR}$, i.e., at $X_{DR}=X_c(q)=\sqrt{g_2/L_{DR}\tau^*q^2}$.  This is the main result as far as how TI affects front propagation. 
The Fourier transformed linear equation is:
%\begin{equation}
%\dot{b}_q=2\frac{\alpha_2}{g_2} Dq^4\tau^*X_{DR}^2 b_q -Dq^2b_q   ~\,.   \label{bqdotD}
%\end{equation} 
%\begin{equation}
%\dot{b}_q=2\alpha_q L_{GT} \tau^* X_{DR}^2 b_q  - Dq^2 b_q -c_q b_q^3  ~\,.   \label{bqdotFront}
%\end{equation}
\begin{equation}
\frac{\partial\chi(\bold{r})}{\partial t}=D\nabla^2\chi(\bold{r}) +\frac{L_{DR}D\tau^* X_{DR}^2}{g_{2}}\nabla^4\chi(\bold{r})    ~\,.     \label{dchiFront}
\end{equation}
%Taking the Fourier transform of the linear Eq.~(\ref{bqdotFront}) gives:
Equation~(\ref{dchiFront}) is very similar to the linearized Kuramoto-Sivashinsky (KS) equation, which also has both $q^2$ and $q^4$ terms~\cite{Desai}.  It is well known that modeling with the KS equation can produce instabilities and interesting pattern formation when the diffusion coefficient is made negative.
   %In both equations an effective negative diffusion coefficient is in the realm of   I do note some important differences;  in the KS equation $D$ is artificially made negative, with no physical justification, in order to get pattern formation.  Also, the sign of the fourth order term is opposite to that of the induction term in Eq.~(\ref{dchiFront}).  It is precisely this term, which at high enough spatial frequencies, that creates an effective ($q$-dependent) diffusion coefficient.  A negative diffusion coefficient creates interesting instabilities and pattern formation.

The KS equation is another important model in the field of nonequilibrium systems and has been successfully utilized to model patterns formed in laminar flame fronts, certain types of Poiseuille flow, trapped ion modes in plasmas, and systems with Eckhaus instabilities such as Rayleigh-B\'enard convection~\cite{Kuramoto1976,SIVASHINSKY1977,MICHELSON1986,Manneville1988,Cross1993,Desai,Cross}. 
Though the SH equation has been used widely as a model, no clear explanation has been provided in the literature for why a diffusion coefficient can be negative.
The TI2 analysis presented here shows precisely how this negative (effective) diffusion coefficient in the KS equation can exist.

The essential physics is now in place for understanding how so many nonequilibrium phenomena may be modeled, i.e., the possibility of tuning user-controlled parameters such as $X_{DR}$ to create an effective diffusion coefficient that is negative, all without violating the SLT. Parameters may be tuned to produce instabilities and patterns at a certain wavevector, $q_{c}$.  These features, along with the enhanced front velocity from Eq.~(\ref{b0dotreal}) are the key elements of the KS model.  The analysis presented here improves upon the KS equation, as it properly accounts for an excitation gap, an essential feature, I believe, for nonequilibrium PTs.
Equation~(\ref{dchiFront}) can be derived from a free entropy functional $\Psi_{2,Q\infty}$ in Eq.~(\ref{Psi2qfunct}) which is constructed in the infinite $Q$ limit.  This provides a good example of the usefulness of the infinite $Q$ analysis, and also motivates the finite $Q$ approach as an improvement which better accounts for fluctuations.

%
%When the KS equation is used to model purely dissipative systems, such as flame front propagation, it suffers from having an unphysical negative diffusion coefficient, thus violating the SLT. 

The analysis presented here is a good example of entropic coupling; This specific solution to the entropic coupling problem shows how real diffusion is overcome by TI to produce a nonequilibrium PT in such a way as to not violate the SLT.  Any patterns that may form from undulations of the front have an entropy cost, but while the pattern is forming, the front as a whole is always moving fast enough to ensure that the total entropy is always increasing.  A similar conclusion is drawn with the reaction-diffusion system discussed in the next subsection.

%The geometric significance of the cutoff term is explained in Ref.~\cite{Cross}, in the context of front propagation.   
%Stability of the front is determined by the sign of $-\nu_{KS}q^2 -\kappa_{KS}q^4$.  

%To produce instability in a front separating two phases, a negative value for the effective diffusion coefficient in the KS equation is required, something which has been difficult to justify on physical grounds.  This justification is now established, since applying a strong enough $X_{DR}$ will always access a mode which will become unstable.  The apparent breakdown of the SLT is taken care of when the DR is included, as discussed above. 
The approach given here can be used to successfully model undulations of a propagating front separating any two phases with a large miscibility gap $\Delta\mu$ ($X_{DR}=-\Delta\mu/T$), ex. fronts in laminar flames, liquid-gas PTs, and autocatalytic chemical reactions.  These types of dissipative systems, as well as any others that can be modeled by TI2, present themselves as strong evidence for the existence of TI.  

%Don't forget potential!

\subsection{Turing Patterns in Chemical Systems} \label{sec:turing}

In Sec.~\ref{sec:front}  I discussed an example of second order TI with a physical interface (front).  In more general terms, a physical interface is not necessary for spontaneous pattern formation.  Turing suggested that a uniform chemical mixture could, under the right reaction conditions, spontaneously self-organize and form patterns~\cite{Turing1952,Cross}.  Nonlinearity is required in the reaction-diffusion equations;  In theory, under the right conditions a bifurcation in the dynamics could be reached and the instability would break symmetry to produce patterns.

Recently, Turing patterns have been observed while observing the chlorite-iodide-malonic acid (CIMA) reaction and the Belousov-Zhabotinsky (BZ) reaction which involves bromide and bromous acid~\cite{Castets1990,Swinney1991,Ouyang1991,Zhabotinsky1991,Swinney1992,Vanag2001}.  For example, the well-studied CIMA reaction is known for producing two-dimensional patterns possessing clear crystal-like symmetry, some with hexagonal patterns, others resembling modulated stripes, as well as mixed states~\cite{Castets1990,Swinney1991,Ouyang1991,Cross1993,Cross}. 
That almost four decades passed between Turing's theoretical paper and the first experimental observations of Turing patterns, shows that in most chemical reactions it is very difficult to produce $X_{DR}$ values large enough to achieve pattern formation.  In the systems which do show patterns, diffusion coefficients must be optimized, presumably bringing the bifurcation point closer to thermodynamic equilibrium and thus easier to access.  Clearly these are not transitions that occur near thermodynamic equilibrium and qualify as nonequilibrium PTs.

For these interesting reactions a thermodynamic analysis is warranted.  If reaching the threshold for this type of nonequilibrium PT is challenging then it makes sense to study how the system will behave both near critical and just below.  I consider a chemical reaction much simpler than the CIMA and BZ reactions.
A second order chemical reaction will suffice to produce the TI effects which will demonstrate the induced pattern formation, so  
%Turing suggested that a uniform chemical mixture could, under the right reaction conditions, spontaneously form patterns~\cite{Turing1952}.
%First order reactions will not suffice since nonlinearity is essential to produce TI.   
%  the model reaction under consideration is the general class of reactions of the form
%\begin{equation}
%(-\nu_A) A+(-\nu_B) B \longleftrightarrow (\nu_C) C  \,,      \label{reaction}
%\end{equation}
%where $\nu_i$ are the stoichiometric coefficients.  Nonlinearity is essential in this analysis so the specific case $\nu_B=-1$, $\nu_C=-2$ (second order), $\nu_D=+1$ will suffice to demonstrate the induced pattern formation.
for the sake of simplicity, the following reaction is considered
\begin{equation}
A+2B \longleftrightarrow C  \,.      \label{reaction2}
\end{equation}
The volume concentration $n_B$ is assumed to be much smaller than $n_A$, making $n_B$ the bottleneck, or gate variable, for the system, i.e., there is plenty of $A$ but a small amount of $B$ that limits the reaction.  
%The general TI approach has been described in Ref.~\cite{Patitsas2014} and for the isothermal case, in Ref.~\cite{Patitsas2015}.  
The extensive thermodynamic variable $N_A=n_A V$ is identified with the dynamical reservoir (DR), i.e., $x_{DR}=N_A$.  When driven away from the equilibrium value $N_{A_0}$ by an amount $\Delta N_A \equiv  a_{DR}$, the relaxation time for the DR is very long so that over the gate timescales, relative changes in $N_A$ are always small.  The thermodynamic force conjugate to $a_{DR}$ is  
\begin{equation}
X_{DR}=-\frac{A}{ T}  \,,      \label{chemforce}
\end{equation}
expressed in terms of the affinity~\cite{mazur,Kondepudi}:
\begin{equation}
A=\mu_A+2\mu_B-\mu_C  \,.      \label{affinity}
\end{equation}
The DR capacitance coefficient, given by $g_{DR}^{-1}=N_A/k_B$, is considered very large in this analysis.

The DR dynamics is given in terms of standard chemical rate theory with the reaction rate $v$ given by
\begin{equation}
v=R_F\left(1-e^{-A/k_B T}   \right)  \,,      \label{diode}
\end{equation}
\begin{equation}
R_F=k_f n_A n_B^2  \,,      \label{nu0}
\end{equation}
where $R_F$ is the forward reaction rate and $k_f$ is the forward rate constant.  For very small affinity Eq.~(\ref{diode}) becomes $v=R_F A/k_B T$ and the dynamics can ordinarily be expressed in terms of Eq.~(\ref{DRdyn}), with the Onsager coefficient given by
%\begin{equation}
%\dot{a}_{DR}=M_{11}Y_{DR}  \,,      \label{DRdot}
%\end{equation}
%where  
\begin{equation}
M_{DR}=\alpha_0 N_B^2 \,,  \label{M11}
\end{equation}
with
\begin{equation}
\alpha_0= \frac{k_f n_A}{V k_B} \,.  \label{alpha0}
\end{equation}
In the limit of very small $n_B$ the bottleneck effect is severe as there is so little of component B available.  In this case the reaction can proceed forwards more rapidly by having some regions gaining in component B at the expense of other nearby regions, i.e., by spontaneously forming undulations in $n_B$.  This type of borrowing and lending of B will be difficult at long wavelengths but rather easy to accommodate at microscopic lengthscales.  This motivates a reaction rate that is proportional to $(\del n_B)^2$ as opposed to simply $n_B^2$.  The coefficient $\alpha_0$ in Eq.~(\ref{M11}) is replaced by $\alpha_0 q^2/q_0^2$ where $q_0^{-1}$ defines a length scale below which the borrowing/lending becomes significant. 

%The relaxation rate for the DR is given by $\tau_{DR}^{-1}=k_r n_B^2$.
%\begin{equation}
%\tau_{DR}^{-1}=k_r n_B^2 \, .    \label{tauDR}
%\end{equation}
When the concentrations vary spatially, Eq.~(\ref{M11}) is modified by replacing $N_B^2$ by the spatial average which is denoted with a tilde as $\widetilde{N_B^2}$.
%The entropy production rate for the DR is 
%\begin{equation}
%\dot{S}_{DR}=M_{11}Y_{DR}^2=\alpha Y_{DR}^2 \widetilde{N_B^2}  \,.      \label{power}
%\end{equation}
%Defining $n_B^*\equiv \widetilde{n_B^2}/\widetilde{n_B}$ the molar entropy production can be expressed as $\sigma_{DR}\equiv \dot{S}_{DR}/\widetilde{N_B}=\alpha Y_{DR}^2 n_B^* V$.
Any patterns that we seek will be expressed by Fourier expansion of the density function for component $B$ as
\begin{equation}
n_B=n_0+ \sum_{q\neq 0}{n_q}\cos(qr-\phi_q)  \,.      \label{fourier}
\end{equation}
Spatial averaging after squaring gives 
\begin{equation}
\widetilde{n_B^2}=n_0^2+\frac{1}{2}\sum_{q\neq 0}{n_q^2} \,.   \label{halfsum}
\end{equation} 
The gate variables are defined as $x_0=n_0 V$, and $b_q\equiv Vn_q$ with $q\neq 0$.  Comparison of Eq.~(\ref{MDR}) with Eq.~(\ref{halfsum}) shows that $\alpha_q=\alpha_0 q^2/2q_0^2$ and $\alpha_2=\alpha_0/2q_0^2$.
%Before obtaining the force conjugate to $a_q$, it is helpful to define a dimensionless quantity $\eta\equiv (n_q/RT)\partial\mu_C/\partial n_q$.  In many chemical systems $\eta\approx 1$ with examples including the ideal classical gas and dilute solutions.  
%The conjugate force is $X_q=-g_{q}b_q$ with $g_{q}^{-1}= Vn_0/k_B$, and the gate variable reaction relaxation rate is $\tau_{q}^{-1}= k_f n_A n_B^2$.  
 
Systems displaying Turing patterns are often referred to as reaction-diffusion systems because of the important role played by diffusion~\cite{Dufiet1992,Desai}.  When the system relaxes to equilibrium via diffusion, $g_{q}= \Delta\mu/\Delta N_B=\zeta\mu_B/n_B V$, where $\zeta$ is a dimensionless parameter typically close to unity.  After incorporating diffusion as well as second order TI and the cut-off term, all the dynamics are well-described by Eqs.~(\ref{b0dotgen}) through to (\ref{psidyn}).  In particular: 
%\begin{equation}
%\dot{b}_{q}=\left( \frac{k_f n_0 n_A A^2 \tau_q^*}{k_B^2 T^2 q_0^2 } -D \right)q^2 b_q -c_q b_q^3~\, . \label{aqdotTuring}
%\end{equation}
\begin{equation}
\dot{b}_{q}=\left(\frac{A^2}{A_{c,q}^2} -1\right)Dq^2 b_q -c_q^2 b_q^3~\, . \label{aqdotTuring}
\end{equation}
Similarly to the KS equation the possibility of an effectively negative diffusion coefficient arises when the affinity is large enough in magnitude.
%A nonequilibrium mode $b_q$ will relax by both chemical reaction and diffusion.  For chemical reaction via reaction~(\ref{reaction2}) the time scale is given by
%\begin{equation}
%\tau_{q,react}^{-1}=\nu_B^2 k_r n_A n_B \, .    \label{tauReact}
%\end{equation}
%Accounting for diffusion with coefficient $D$, gives the net rate as
%\begin{equation}
%\tau_{q,0}^{-1}=\tau_{q,react}^{-1}+\tau_{q,diff}^{-1}=4 k_r n_A n_B +Dq^2  \, .    \label{tauq}
%\end{equation}
%Thermodynamic induction will affect the the mean concentrations via $b_0$ as well as the undulating modes $n_q$.
%This type of induction on the mean concentrations $n_A$, $n_B$ has essentially been treated already in Ref.~\cite{Patitsas2015} Sect. 6, where one can replace the bias applied across parallel plates with the affinity $A$.  Though this effect will not create a Turing pattern, second order TI on $b_q$ with $q\neq 0$ does have the potential to do so.
%Making use of the general results from Section~\ref{sec:gentheory}, the dynamics for $b_q$, using Eq.~(\ref{aqdot}), are given by
%\begin{equation}
%\dot{a}_{q}=\alpha_q X_{DR}^2 L_{qq}\tau_q^* a_q  -L_{qq}g_{q}a_q~\, . \label{dum}
%\end{equation}
The critical condition for the affinity is
%, using Eq.~(\ref{Xcrit}), is
%\begin{equation}
%X_{DR,crit}(q)= \sqrt{\frac{g_{q}}{\alpha_q \tau_q^*}}        ~\, . \label{Xcrit}
%\end{equation}
%\begin{equation}
%-A_{crit}(q)/T= \sqrt{\frac{2k_B V k_B}{Vn_0 k_r n_A \tau_q^*}}        
%\end{equation}
%\begin{equation}
%|A_{c}(q)|=  k_B T\sqrt{\frac{1+\frac{D }{k_f n_0 n_A}q^2 }{{k_f n_0 n_A \tau_q^*}}}         ~\, . \label{Acrit}
%\end{equation}
%\begin{equation}
%|A_{c}|=  k_B T\sqrt{\frac{Dq_0^2}{{k_f n_0 n_A \tau_q^*}}}         ~\, . \label{Acrit}
%\end{equation}
%\begin{equation}
%|A_{c}|=  \sqrt{\frac{q_0^2\zeta k_B^2 T^2}{{q^2 k_f n_0 n_A \tau_q^*}}}         ~\, . \label{Acrit}
%\end{equation}
%\begin{equation}
%|A_{c}|=  \sqrt{\frac{q_0^2 g_q k_B N_B T^2}{{q^2 k_f n_0 n_A \tau_q^*}}}         ~\, . \label{Acrit}
%\end{equation}
\begin{equation}
|A_{c,q}|=  \sqrt{\frac{q_0^2 g_q  T^2}{{q^2 \alpha_0\tau_q^*}}}         ~\, . \label{Acrit}
\end{equation}
%Note: this implies that even small A will produce patterns at large q.  We need a cutoff.
%\begin{equation}
%A_{crit}(q)= -k_B T\sqrt{\frac{2 n_0}{R_F \tau_q^*}}        ~\, . \label{Acrit2}
%\end{equation}
%\begin{equation}
%|A_{crit}(q)|= k_B T\sqrt{\frac{\tau_{q,react}}{2 \tau_q^*}}        ~\, . \label{Acrit3}
%\end{equation}
%its a condition on A2 actually
%\begin{equation}
%A_{crit}^2(q)= (k_B T)^2\frac{\tau_{q,react}}{2 \tau_q^*}        ~\, . \label{Acrit4}
%\end{equation}
Thus, in an experiment where concentration levels for A or C are pushed from equilibrium far enough, the system will become unstable and Turing pattern formation could occur.  Equation~(\ref{Acrit}) defines the dispersion relation and shows clearly the existence of a threshold;  Control parameters such as reactant concentrations are adjusted until $A$ is away from equilibrium by a sufficient amount.  This is indeed observed in the reported literature.  For example in one report on the CIMA reaction, a clear threshold at a specific malonic acid concentration was observed as the system transitioned from the normal state to a striped pattern~\cite{Swinney1992}.  Even though the CIMA reaction has a good deal more complexity than Eq.~(\ref{reaction2}) and there exists more than one bifurcation point, the essential feature of a threshold for a nonequilibrium PT has been established.  Note that the analysis presented here does not predict the precise type of patterns that are observed in CIMA and BZ reactions;  These reactions are more complex than Eq.~(\ref{reaction2}).  Reported analysis of the stability equations for the CIMA reaction have been shown to predict specific patterns~\cite{Pismen1980,Cross1993,Desai}.  Fluctuations should play an important role in the specifics of pattern formation.  Indeed, the long range order analysis presented below in Sec.~\ref{fluctmany} shows that pattern formation could be more robust in lower-dimensional systems. Thus, with more work, TI2 could explain multiple bifurcation points and complex phase diagrams. 

%The preliminary analysis presented here can be extended;  For example after a stripe phase is formed, increasing $X_{DR}$ further could access a second nonequilibrium PT where the stripes are destabilized and break up into dots.  Indeed, the long range order analysis presented below in Sec.~\ref{fluctmany} shows that pattern formation of the lower-dimension dots may be more robust than for striped patterns.  Thus, with more work, TI2 could explain multiple bifurcation points and complex phase diagrams.

The dynamics of Eq.~(\ref{aqdotTuring}) can be derived from a free entropy functional $\Psi_{2,Q\infty}$ in Eq.~(\ref{Psi2qfunct}) which is constructed in the infinite $Q$ limit, thus providing another application of this useful functional.  It is interesting to go beyond this approach and include effects of nonequilibrium fluctuations.
For example, my analysis shows that just below critical there should be large fluctuations that could be interpreted as chemical turbulence.  It is interesting that the CIMA reaction does indeed show a transition between striped patterns and chemical turbulence under certain conditions~\cite{Swinney1991}.  This might be the result of frustration between hexagons, stripes and rhombic patterns, but it might also be the result of thermodynamic fluctuation just below the instability point.  If so then the dynamics would be described by the following equation for the fluctuations in $b_q$: 
\begin{equation}
\dot{u}_q=2\left(\frac{X_{DR}^2}{X_{c}^2(q)}-1\right) \frac{u_q}{\tau_q}  +2\left(\frac{X_{DR}^2}{X_{c}^2(q)}\right) \frac{\langle b_q^2\rangle_0}{\tau_q}-2Dq^2 u_q -2c_q^2 u_q^2    ~\,.   \label{uqdotTuring}
\end{equation}
Large thermodynamic fluctuations below critical might also explain how for the case of the BZ reaction, reports make mention of a long-lived and complicated transient state before a sustained pattern is produced~\cite{Winfree1983,Cross}.

The long range order analysis presented below in Sec.~\ref{fluctmany} shows that nonequilibrium fluctuations may make pattern formation difficult in 3 dimensions, and likely easier in 2 dimensional systems.  Given the typical design of reaction cells in the literature~\cite{Winfree1983,Castets1990,Swinney1991,Ouyang1991,Swinney1992,Vanag2001} as a gel contained between two closely spaced plates, it makes sense that pattern formation may be difficult for $q$ values larger than the inverse of the plate spacing (typically 0.2 mm for the CIMA reaction~\cite{Swinney1992}).

It is worth pointing out the similarities between this type of Turing pattern formation in reaction-diffusion systems and what occurs in Langevin B systems immediately after a quench through an equilibrium PT, with initial conditions being a completely uniform system.  Well before formation of any clear fronts and boundaries between two distinct phases, density undulations at certain wavevectors would be required to break symmetry.  The equations presented here predict that these undulations would occur at larger spatial frequencies, at least for short times.  These undulations would after long times develop into fronts separating phases. 

Finally, I note that I have now produced two good examples with nonequilibrium PTs and in both cases the PTs begin first at high spatial frequencies for modest $X_{DR}$.  More examples are needed but this behaviour at high spatial frequencies is an indicator of universal behaviour~\cite{SethnaBook}. 
In these two examples the dynamics are generally quite slow and inertial effects are not prominent.  Future TI work on faster systems such as ultrasonic propagation and second sound would require more development.  The recent formulation of Extended Irreversible Thermodynamics is well-suited for studying such fast phenomena~\cite{Lebon1988}.  Incorporating the DR into this formulation should allow for future TI studies of such fast systems. 

%\subsubsection{Thermodynamic Potentials}
%
%The appropriate potentials at this level of analysis are the infinite $Q$ potentials $\Phi_{2,\infty Q}(b)$ and $\Psi_{2,\infty Q}(b)$.  
%For $\Phi_{2,\infty Q}(b)$, Eq.~(\ref{Phi2}) is easily modified to include a summation over states $q$:
%% Note replace tauGT with 1/Dq2
%\begin{equation}
%\Phi_{2,\infty Q}(b)=L_{DR}X_{DR}^2+X_{DR}^2\alpha b^2 -\sum_{q\neq 0}{\lambda_q \left(-2\alpha \left(X_{DR}^2 -X_{c,q}^2\right) L_{GT}g_{GT}\tau^* b^2 +c_q^2 g_{GT} b^4 \right)}   ~\, , \label{Phi2Turing}
%\end{equation}  
%\begin{equation}
%\Phi_{2,\infty Q}(b)=\sum_{q\neq 0}{X_{DR}^2\alpha b_q^2} -\sum_{q\neq 0}{\lambda_q \left[-2\alpha \left(X_{DR}^2 -X_{c,q}^2\right) L_{GT}g_{GT}\tau^* b^2 +c_q^2 g_{GT} b^4 \right]}   ~\, , \label{Phi2Turing}
%\end{equation} 
%\begin{equation}
%\Phi_{2,\infty Q}(b)=\sum_{q\neq 0}{\left[\frac{g_q}{2\tau^*}\frac{A^2}{A_c^2}b_q^2 +\lambda_q g_q\left(\frac{A^2}{A_c^2}-1\right)Dq^2 b_q^2 -\lambda_q g_qc_q^2 b^4 \right]}   ~\, , \label{Phi2Turing}
%\end{equation}
%where $\lambda_q$ above critical is     
%\begin{equation}
%\lambda_{q}= \frac{\tau_{GT}}{2\tau^*}\frac{X_{DR}^2}{X_{DR}^2-X_{c,q}^2}   ~\, . \,\,\,\,\, \text{(above critical)} \label{lambdaq}
%\end{equation}
%\begin{equation}
%\lambda_{q}= \frac{1}{2\tau^* Dq^2}\frac{A^2}{A^2-A_{c,q}^2}   ~\, . \,\,\,\,\, \text{(above critical)} \label{lambdaq}
%\end{equation}
%
%psi2 is available, but wait below for psi2 of u?

%\vspace{2in}

\subsection{Fluctuations with Many Modes and Long Range Order}\label{fluctmany}

Any patterns that may be formed near and/or above a critical point will be determined by a sum over Fourier modes $q$, and such patterns would be examples of order-by-induction.  Some level of long range order would be required to create a discernible pattern and any function used to describe long range order should also involve a (weighted) summation over $q$ of plane waves.  The dispersion relation $X_c(q)$ will play an important role in the weighting.   For example, if all waves $e^{iqx}$ are weighted equally then there will be no long range order and one would not expect pattern formation.  Instead one may see an increase in fluctuations at all length scales, a result resembling hydrodynamic turbulence.  Such an effect, termed chemical turbulence, has been observed in reaction-diffusion systems~\cite{Ouyang1991,Mecke2015}.  It may indeed be the case that in systems exhibiting chemical turbulence, the dispersion relation is very flat.
In contrast, if the dispersion relation has a prominent minimum at $q_{min}$ then one might expect a clear pattern resembling a standing wave with wavevector $q_{min}$.  However, fluctuations may play an important role in disrupting pattern formation. 

For equilibrium PTs it is well known that thermal fluctuations can disrupt long range order.  The Mermin-Wagner theorem dictates that long range order is lost in one and two dimensional systems because of fluctuations at long wavelengths~\cite{Wagner1966,Gelfert2001}.  To date the only way around this theorem is the Kosterlitz-Thouless PT which involves ordering in topological defects such as magnetic vortices in two dimensions~\cite{Kosterlitz_1973}. 
Many of the observations of pattern formation in nonequilibrium systems are essentially two dimensional systems.  These systems may also be interesting exceptions to the Mermin-Wagner theorem.  Indeed it may be easier to achieve long range order in one and two dimensions near nonequilibrium  PTs than it is for equilibrium PTs. For equilibrium PTs long range order is determined using the Boltzmann factor $\exp{(-H/k_B T)}$ in the statistical weighting.  The Hamiltonian $H$ typically goes as $q^2$ as it does for systems with spin-spin interactions.  The $q^2$ dependence makes the weighting factor soft for long wavelength (Goldstone) modes~\cite{Bergerson}.  In contrast the weighting factor for TI2 is $\exp{(\Psi_2(u))}$.  The extent of fluctuations is controlled by the term $-\frac{1}{2}g_4 u^2$ which does not scale as $q^2$ and is not expected to depend on $q$, as the entropy cost of an undulation depends on the amplitude of the undulation, not the slope.  Thus the weighting factor for TI2 remains stiff at large length scales.  This means that one does not expect long wavelength Goldstone modes to disrupt long range order.

For the case where the dispersion relation $X_c(q)$ has a minimum at finite $q_{min}$ then for large $Q$ the PT occurs when $X_{DR}$ is raised up to $X_c(q_{min})$, with the expectation of a clear pattern and long range order.  When $X_{DR}$ is raised slightly higher, modes with $q$ values over the range from $q_1$ to $q_2$, where $X_c(q_1)=X_{DR}$ and $X_c(q_2)=X_{DR}$, will be excited.  These excitations are the Goldstone modes for this case.  Following the example for equilibrium PTs, Goldstone modes are excitations on top of the (nonequilibrium) PT that have a small cost in $\Psi_2$, or the nonequilibrium free energy, $\mathpzc{F}$.  If, for the sake of simplicity, each excited mode is weighted equally, the sum: $\langle\psi\rangle=\sum_q\cos(\bf{q}\cdot\bf{r})$ can show effects of the disruption of long range order.  These sums are similar to the ones used in elementary optics calculations for coherence in diffracting systems.
For example, in one dimension $\langle\psi\rangle=q_2 \text{sinc}(q_2 x)-q_1 \text{sinc}(q_1 x)$.  When $\Delta q=q_2-q_1$ is small, $\langle\psi(x)\rangle=2\Delta q\text{sinc}(\Delta q x)\cos(q_{min}x)$, and one identifies the correlation function as the envelope of $\langle\psi\rangle$, here $2\Delta q\text{sinc}(\Delta q x)$.  As expected then, as $X_{DR}$ is raised higher, $\Delta q$ grows and the Goldstone modes decrease long range order. 
Thus if the quality factor is very high and if $X_{DR}$ is finely tuned enough then formation of a clear pattern is guaranteed.  In some systems $q_{min}$ may be very large and the pattern not easily observed at small length scales.   

For a more general analysis of long range order the correlation function $\langle \psi(\bf{r})\psi(0)\rangle$ should be investigated.  Taking the Fourier transform and ignoring a prefactor gives
\begin{equation}
K(\textbf{r})\equiv\sum_q e^{i\textbf{q}\cdot\textbf{r}}\langle \text{u}_{\text{q}}\rangle    ~\,.   \label{corr}
\end{equation}
The average $\langle u_q\rangle$ is taken with the weighting factor $\exp{(\Psi_2(u_q))}$.  The result has already been calculated as the stationary state from Eq.~(\ref{uquad}), re-expressed as
\begin{equation}
\langle u_q\rangle=\langle b_q^2\rangle_0+2Q^2\langle b_q^2\rangle_0\left[z_q^2-1+\sqrt{(z_q^2-1)^2+z_q^2 /Q^2}\right]    \,,   \label{uqavg}
\end{equation}
where $z_q=X_{DR}/X_c(q)$.  The function describing long range order would be the modulus of $K(\textbf{r})$.  When $X_{DR}$ is below the excitation gap, $K$ is small and then becomes substantial as the excitation threshold is reached.  
%Well above threshold, $K$ grows as $X_{DR}^2$. 
In the two important examples discussed in Sec.~\ref{sec:front} and Sec.~\ref{sec:turing}, the induction was derived from first principles and in both cases $\alpha_q$ goes as $q^2$ and the dispersion relation $X_c(q)$ goes as $q^{-1}$.  This would imply gapless excitation at infinite $q$.  However, at microscopic length scales induction is expected to taper off, and there is a physical cutoff $q_{cut}$ for the wave vector.  For example, in a chemical reaction the mean free path would likely dictate the value of $q_{cut}$.  The excitation gap is then $X_c(q_{cut})$ and this would likely be much smaller than the excitation gap in a system with a flat dispersion relation.  Though, systems with $\alpha_q$ proportional to $q^2$ will allow easier excitation of nonequilibrium PTs, these transitions are more difficult to discover and study with the very small pattern period.

If $X_{DR}$ is raised further above $X_c(q_{cut})$ the period of the pattern will increase and should become easier to observe.  For such $X_{DR}$ the critical wavevector is $q_c<q_{cut}$ such that $X_c(q_{c})=X_{DR}$. In the calculation for $K(\textbf{r})$, $q$ values with magnitudes between $q_1$ and $q_{cut}$ will make the largest contributions.
For large $q$ the expansion $\sqrt{1+q^2b_q^2}\approx 1+\frac{1}{2}q^2b_q^2$ used in Sec.~\ref{sec:front} for the eikonal will break down.  This produces a natural cutoff at large $q$ since $\sqrt{1+q^2b_q^2}\approx qb_q$ means that TI2 becomes TI1 which leads to a stationary state $b_{q,ss}=1/q$, and $u_{q,ss}=1/q^2$.  This result represents a type of universality since any nonequilibrium front will show this $b_{q,ss}=1/q$ behaviour for $q>q_{c}$.
The critical wavenumber $q_c$ will set the periodicity of any patterns and one would identify this mode with the PT.  Technically, all wavenumbers larger than $q_c$ are involved in individual PTs.  Since excitation of these modes tend to suppress long range order of the mode at $q_c$, they may be thought of as Goldstone modes.  For these types of nonequilibrium PTs, Goldstone modes present themselves at high spatial frequencies and may be generally characterized as turbulence.
%  A complementary picture is presented for nonequilibrium phase transitions    It is   

When $u_q=1/q^2$ the sum in Eq.~(\ref{corr}) allows easy calculation of $K(\textbf{r})=r^{2-d}\cos{(q_cr)}$.  In one and two dimensions, the long range order is strong, and somewhat marginal in three dimensions.  This trend is opposite to what happens in equilibrium PTs where long range order is disrupted by Goldstone modes at low dimensions.

Finally, I note that these results favour the possibility of fluctuations at large $q$ being enhanced by TI.  If so, then the physics at the nanometer scale becomes even more interesting;  At very small length scales systems that are profoundly influenced by thermal and quantum fluctuations, may also be influenced by these types of nonequilibrium fluctuations.

\subsection{Nonequilibrium Thermodynamic Identity}

The main result presented as $\mathpzc{S}_{NE}$ being maximized by stationary states, may be expressed as $d\mathpzc{S}_{NE}=0$.  For equilibrium thermodynamics, the total entropy $S_T$ is maximized which leads to the thermodynamic identity: $
dU=TdS-PdV+\sum_i\mu_i dN_i$, where $S$ refers to the system of interest that is removed from the thermal reservoir, particle reservoirs, etc..  

The generalized entropy $\mathpzc{S}_{NE}$ accounts for the gate which includes traditional thermodynamic reservoirs.  The next step is to take the gate and separate away the system of interest from such reservoirs including but not limited to thermal, volume, and particle reservoirs.
%The distinction between SI and gate, in particular between $S_{GT}$ and $S_{SI}$, now becomes explicit.
The variables $x_{GT}$ and $b$ are adapted to $x_j$, $b_j$ respectively, where $x_j$ can be $U$, $V$, $N_i$, etc.  This accounts for the Onsager coefficient $M_{DR}$ depending  on internal energy, volume, particle numbers, etc. so that $\gamma_j=\partial M_{DR}/\partial x_j$.
Equations~(\ref{SNE2}) and (\ref{Xi}) are modified to account for more than one gate variable:
\begin{equation}
\mathpzc{S}_{NE}= S_{GT}+\sum_j  \Xi_{0,j}+\sum_j k_B\beta_{1,j}b_j +\sum_j k_B\beta_{2,j}u_j  \,.  \label{SNEfull} 
\end{equation}
Separation of the system of interest away from the traditional reservoirs is straightforward, though care must be taken to distinguish the standard variables such as $U$, $V$ $N_i$, from the mode variable $b_q$ that become excited in TI2.  At this point I will explicitly treat the three types reservoirs, thermal, volume, and particle, where it is understood that there are least two components present (ex. 2 phases to form a front, or 3 components for chemical reaction, as in the above examples) which  allows for one variable with Fourier modes $b_q$ and $u_q$:
%\begin{equation}
%\mathpzc{S}_{NE}=S-\frac{U}{T} -\frac{PV}{T}+\frac{\mu N}{T}+S'+k_B\beta_{1,U}(U-U_{eq})+k_B\beta_{1,V}(V-V_{eq})+k_B\beta_{1,N}(N-N_{eq}) +k_B\beta_{2,U}u_U+k_B\beta_{2,V}u_V+k_B\beta_{2,N}u_N  \,.  \label{SNEres}
%\end{equation}
\begin{eqnarray}
\mathpzc{S}_{NE}&=&S-\frac{U}{T} -\frac{PV}{T}+\frac{1}{T}\sum_i{\mu_i N_i}+S' \nonumber \\ 
&&+k_B\beta_{1,U}(U-U_{eq})+k_B\beta_{1,V}(V-V_{eq})+k_B\sum_i{\beta_{1,N_i}(N_i-N_{i,eq})} \nonumber \\
&&+k_B\beta_{2,U}u_U+k_B\beta_{2,V}u_V+k_B\sum_i \beta_{2,N_i}u_{N_i}  \nonumber \\
&&+k_B\sum_{q\neq 0}\beta_{2,q}u_q  \,,  \label{SNEres}
\end{eqnarray}
where $S$ is the entropy of the system of interest, and $S'$ is constant.  When applying the condition $d\mathpzc{S}_{NE}=0$ it is understood that $U$, $V$, $N$, $u_U$, $u_V$, and $u_N$ are varied independently while $T$, $P$, $\mu$, $\beta_{1,U}$, etc. are held constant.  The result is a revised version of the thermodynamic identity:
\begin{eqnarray}
dU&=&TdS-PdV+\mu dN \nonumber \\ \nonumber
&&+k_BT\beta_{1,U}dU+k_BT\beta_{1,V}dV+k_BT\sum_i\beta_{1,N_i}dN_i \\  \nonumber
&&+k_BT\beta_{2,U}du_U+k_BT\beta_{2,V}du_V+k_BT\sum_i\beta_{2,N_i}du_{N_i} \\
&&+k_BT\sum_{q\neq 0}\beta_{2,q}du_q  \,.  \label{dU}
\end{eqnarray}

Equation~(\ref{dU}) is illuminating as to the influence of the DR on the system of interest.  When $X_{DR}$ is held constant by some external agent then the dynamical reservoir has the same type of effect on the system of interest as the traditional reservoirs.  For example, the effect of $M_{DR}$ depending on $V$ gives an effective pressure $P_{eff}=P-k_BT\beta_{1,V}$.  Thus, the classical thermodynamic variables can be altered when the DR is well away from equilibrium.

Just as working under conditions of constant temperature, pressure, and chemical potential is often more convenient, so is likely the case for constant $\beta_{1,U}$, etc.  Legendre transformations for the nonequilibrium variables are just as straightforward as for the equilibrium variables.  For example a generalized Gibbs potential may be defined as
\begin{equation}
\mathpzc{G}\equiv U+PV-ST-k_BT\beta_{1,U}U -k_BT\beta_{1,V}V-k_BT\beta_{2,U}u_U-k_BT\beta_{2,V}u_V    \,.  \label{G}
\end{equation}
Implementing the thermodynamic identity gives for the differential:
%\begin{equation}
%d\mathpzc{G}=dU+PdV+VdP-SdT-TdS-k_BT\beta_{1,U}dU-k_BUd(T\beta_{1,U})-k_BT\beta_{1,V}dV-k_BVd(T\beta_{1,V})-k_BT\beta_{2,U}du_U-k_Bu_U d(T\beta_{2,U}u)-k_BT\beta_{2,V}du_V-k_Bu_Vd(\beta_{2,V})    \,.  \label{dG}
%\end{equation}
\begin{eqnarray}
d\mathpzc{G}&=&\sum_i\mu_i dN_i+VdP-SdT+k_BT\sum_i\beta_{1,N_i}dN_i+k_BT\sum_i\beta_{2,N_i}du_{N_i} \nonumber \\
&& +k_BT\sum_{q\neq 0}\beta_{2,q}du_q   \nonumber \\
&&-k_BUd(T\beta_{1,U})-k_BVd(T\beta_{1,V})-k_Bu_U d(T\beta_{2,U}u)-k_Bu_Vd(T\beta_{2,V})    \,.  \label{dG}
\end{eqnarray}
Under conditions of constant $T$, $P$, $u_{N_i}$, $\beta_{1,U}$, $\beta_{1,V}$, $\beta_{2,U}$, $\beta_{2,V}$, and $u_q$, $d\mathpzc{G}=\sum_i\mu_{i,eff}dN_i$, where $\mu_{i,eff}=\mu_i+k_BT\beta_{1,N_i}$.  

New Maxwell relations can be derived from Eq.~(\ref{dG}).  For example $\partial\mathpzc{G}/\partial\beta_{1,U}=-k_BTU$ and $\partial\mathpzc{G}/\partial\beta_{1,V}=-k_BTV$ and equating the crossing second order derivatives merely results in a consistency check for $\partial^2 M_{DR}/\partial U\partial V=\partial^2 M_{DR}/\partial V\partial U$.
Another Maxwell relation is created by crossing $P$ and $\beta_{1,V}$, giving
\begin{equation}
\frac{\partial V}{\partial\beta_{1,V}}=-k_BT\frac{\partial V}{\partial P}    \,.  \label{Max}
\end{equation}
It is understood that the isothermal compressibility on the right hand side of Eq.~(\ref{Max}) is taken at constant $\beta_{1,V}$ and is not the same as the traditional isothermal compressibility.
% since $\partial^2\mathpzc{G}/\partial\beta_{1,V}\partial\beta_{1,U}=-k_BT\partial U/\partial\beta_{1,U}$ 

Establishment of the nonequilibrium thermodynamic identity, Eq.~(\ref{dU}) shows that TI can be incorporated into the existing thermodynamic framework by making use of the DR.  
%The same type of approach can also be used to establish a nonequilibrium partition function as will be shown in Sec:~\ref{sec:canon}.

\subsection{Entropic Coupling Problem With Many Modes} \label{Sec:EntCoup2}

%Equation~(\ref{SGTquad}), along with the nonequilibrium Le Chatelier's principle, up to 2nd order, solves the entropy coupling problem.  The SI can achieve a state that can be referred to as patterned, or self-organized, or self-assembled, by having its entropy lowered relative to its equilibrium, and during the entire time of this self-assembly, the second law of thermodynamics is never violated.  The result hold for first order TI and for second order, both below and above the critical bifurcation point.  The DR plays a key role as that system which always increases it entropy by at least as much as the decrease in the entropy of the SI.  I stress the dynamical nature; If the DR returns to equilibrium, the entropy budget $\Delta S_{GT,ss}$ goes to zero.  

Returning to the important entropic coupling problem, most of the interesting nonequilibrium systems that exhibit self-organization or pattern formations would occur with many modes $q$ being excited.  A variety of models exist in the literature that are capable of exhibiting such pattern formation and self-organization~\cite{Desai,Cross}.
The TI analysis presented here gives new insights into pattern formation and the key to gaining these insights is incorporation of the DR into the analysis.  With the DR properly accounted for, one sees that the spontaneous formation of patterns in the gate do not violate the second law of thermodynamics and that no patterns can form if the DR is at equilibrium, or even below the excitation gap.  Moreover, the creation of such patterns coincides with an \textit{increase} in overall entropy production.  In fact, the total rate of entropy production is maximized.
This leads to the somewhat dualistic physical interpretation; 1) that a large $X_{DR}$ above the excitation gap is what drives the pattern formation or 2) that the patterns form as the gate facilitates the DR to create entropy at the greatest rate possible.

Since each mode $q$ acts independently and always adds a positive term to the total entropy production, there is no limit in principle to how many modes that can become excited, as long as the $\alpha_q$ coefficients are large enough.  With the assumption that all such modes are stationary the entropy change of the gate due to pattern formation is given by
\begin{equation}
\Delta S_{GT,ss}=-2k_B \sum_q Q_q^2\left[z_q^2-1+\sqrt{(z_q^2-1)^2+z_q^2 /Q_q^2}\right]    \,.   \label{SGTquad2}
\end{equation} 
With many modes excited this entropy reduction can be substantial.  If one associates this entropy budget with information storage, then it would seem that the more modes excited the more information that can be stored.  If it assumed that this information would be stored in any pattern that forms then the situation is not so simple;  From the results of Sec.~\ref{fluctmany} more excited modes may actually disrupt pattern formation, because of nonequilibrium fluctuations of Goldstone modes.  There would seem to be a trade-off as far as the benefit of more excited modes goes.  Clearly, more study is warranted on the topic of information storage in nonequilibrium systems.   %~\cite{Lebon1988}

\section{Conclusions}  \label{sec:conc}

%START THIS OVER:

%Discussion of how X-X model uses an ru term that creates instability but when applied to certain systems it is not clear what physics is behind a positive r model.  Here I demonstrate how TI provides the essential physics for creating these important instabilities in dissipative systems.  -for example diffusion, chemical reactions.
%
%major points:
%TI extended to 2nd order:
%thermo of bifurcation
%noneq fluct diss theorem
%solved entropy coupling problem
%conductance over energetics
%nonequilibrium phase transitions well away from equil
%established connection to models and experiments showing spontaneous symmetry breaking
%copious evidence, in contrast to 1st order TI
%maximum entropy production, minimum F
%noneq canonical dist
%4 laws on noneq thermo
%0) stat state plays role of equilibrium state
%1) Sn S Psi (additivity)
%2) max SNE (variational principle)
%3) Psi zero at equil 

With the general thermodynamics presented here, it is the right time now for the research community to discuss the establishment of a new set of physical principles suitable for governing nonequilibrium systems.  A new thermodynamic potential, the generalized entropy, has been constructed such that it is maximized in nonequilibrium situations when the traditional entropy is not maximized.  The generalized entropy is maximized when  the gate is in a stationary state.  Moreover when the entire system is returned to equilibrium the generalized entropy becomes the traditional entropy.  These results have been summarized in a form very much analogous to the laws of equilibrium thermodynamics.

The key to arriving at these results was in identifying the dynamical reservoir and separating this away from the gate, in much the same way as traditional reservoirs are treated.
This separation also answers the historically important entropic coupling problem.  It is now clear that the second law of thermodynamics is never violated even when patterns self-organize in the gate.  During periods where the entropy of the gate decreases, the dynamical reservoir creates entropy at a greater rate, ensuring that the total rate of entropy production is always positive. In two important examples, front propagation and Turing pattern formation, the dynamical reservoir is simply the zero wavevector mode.  In these examples, pattern formation with long range order has been shown to be possible, although the disruptive influence of Goldstone modes has been pointed out.  In contrast to the case of equilibrium second order phase transitions, long range order in patterns is more robust at lower dimensions.

At the heart of the results presented here is thermodynamic induction.  Developing thermodynamic induction up to second order now allows one to explain spontaneous symmetry breaking and phase transitions in the nonequilibrium realm.  Real space examples with many wavevector modes have been
discussed here and second order thermodynamic induction in these examples leads to pattern formation at nonequilibrium phase transitions.  Thermodynamic induction, as a first principles theory, works well at producing the equations and models such as the Swift-Hohenberg, Edwards-Wilkinson, and Kuramoto-Sivashinsky equations and that therefore TI provides a good description of all the systems successfully modeled and reported so far in the literature.  Logically then, the large number of observations of nonequilibrium phenomena that have been successfully modeled with these equations, becomes evidence for the existence of thermodynamic induction.
For example, the production and observations of patterns in laminar flame fronts is to be considered as evidence for second order thermodynamic induction since the first principles results presented here for second order thermodynamic induction predict the Kuramoto-Sivashinsky equation and the Kuramoto-Sivashinsky equation has been developed and used to model the behaviour of these flame fronts.

The approach presented here for obtaining dynamical equations emphasizes conductance, not energetics.  An approach based on energetics, such as a Hamiltonian-based approach, was destined to be difficult, as evidenced by a lack of a solution to the entropic coupling problem since the second law of thermodynamics was established.  It is my hope that this work stimulates more study in the research community of the dynamics of variables coupled to each other via a conductance, or kinetic coefficient.

\section{Acknowledgments}

I thank Cathy J. Meyer for her assistance in editing the manuscript.

\section{Appendix A}

Here is treated the case of two random walk variables, $a$, and $b$, governed by a multiplicity-type statistical weighting factor $\Omega[a,b]\equiv\exp(-1/2 q_1 a^2 -1/2 q_2 b^2)$.  Variable $b$ will have a constant step length $l_2$ while the step length for $a$ is variable, of the form $l(b)$.  If the two variables begin a time step at $(a,b)$ then the probability of both variables moving to the right (increasing) is proportional to $\Omega[a+l(b+l_2),b+l_2]$.  The 4 possible outcomes with probabilities are:
\begin{equation}
p_{rr}=c \Omega[a+l(b+l_2),b+l_2]    \,,     \label{prr}
\end{equation}
\begin{equation}
p_{lr}=c \Omega[a-l(b+l_2),b+l_2]    \,,     \label{plr}
\end{equation}
\begin{equation}
p_{rl}=c \Omega[a+l(b-l_2),b-l_2]    \,,     \label{prl}
\end{equation} 
and
\begin{equation}
p_{ll}=c \Omega[a-l(b-l_2),b-l_2]    \,.     \label{pll}
\end{equation} 
The normalization constant $c=1/4\Omega[a,b]$ when the step sizes are assumed small.  The mean displacement for the single step is then
\begin{equation}
\langle\Delta a\rangle= p_{rr}l(b+l_2) -p_{lr}l(b+l_2)+p_{rl}l(b-l_2) -p_{ll}l(b-l_2)  \,,     \label{Deltaa}
\end{equation} 
and
\begin{equation}
\langle\Delta b\rangle= \left[p_{rr} +p_{lr}-p_{rl}-p_{ll}\right]l_2  \,.     \label{Deltab}
\end{equation} 
Making use of the form of $\Omega$ gives
\begin{eqnarray}
\frac{\langle\Delta b\rangle}{l_2}=\frac{1}{2}\cosh{\left[q_1 a l(b+l_2)\right]}e^{-1/2 q_1 l^2(b+l_2)}e^{-q_2 b l_2} \\
            -\frac{1}{2}\cosh{\left[q_1 a l(b-l_2)\right]}e^{-1/2 q_1 l^2(b+l_2)}e^{q_2 b l_2}  \,.     
\end{eqnarray} 
For TI1, $l(b)$ is specified as $l^2=l_0^2+\delta b$, while for TI2, $l^2=l_0^2+\beta b^2$. After some algebra, again, assuming small step sizes:
\begin{equation}
\frac{\langle\Delta b\rangle}{l_2^2}=-q_2 b+\frac{q_1\delta}{2}(q_1 a^2 -1) \,. \,\,\,\,\,\,\,\,\,\,\,\,\, \text{(1st order)}        
\end{equation} 
\begin{equation}
\frac{\langle\Delta b\rangle}{l_2^2}=-q_2 b+q_1 \beta(q_1 a^2 -1)b \,. \,\,\,\,\,\,\,\,\,\,\,\,\, \text{(2nd order)}       \label{AppA} 
\end{equation}
Similarly one finds:
\begin{equation}
\langle\Delta a\rangle=-q_1 a [l(b)]^2 \,.         
\end{equation}

%To calculate $c$:
%\begin{equation}
%\left[p_{rr} +p_{lr}+p_{rl}+p_{ll}\right]/c\Omega[a,b]=e^{-q_1 al_1(b+l_2)-1/2 q_1 l_1(b+l_2)^2}e^{-g_2 b l_2-1/2 g_2 l_2^2} +e^{q_1 al_1(b+l_2)-1/2 q_1 l_1(b+l_2)^2}e^{-g_2 b l_2-1/2 g_2 l_2^2}  \,.     
%\end{equation}
%\begin{eqnarray}
%\left[p_{rr} +p_{lr}+p_{rl}+p_{ll}\right]/c\Omega[a,b]=e^{-q_1 al_1(b+l_2)-1/2 q_1 l_1(b+l_2)^2}e^{-q_2 b l_2-1/2 q_2 l_2^2} \\
%+e^{q_1 al_1(b+l_2)-1/2 q_1 l_1(b+l_2)^2}e^{-q_2 b l_2-1/2 q_2 l_2^2} \\
%+e^{-q_1 al_1(b-l_2)-1/2 q_1 l_1(b-l_2)^2}e^{q_2 b l_2-1/2 q_2 l_2^2} \\
%+e^{q_1 al_1(b-l_2)-1/2 q_1 l_1(b-l_2)^2}e^{q_2 b l_2-1/2 q_2 l_2^2}      
%\end{eqnarray}
%\begin{eqnarray}
%e^{1/2 q_2 l_2^2}\left[p_{rr} +p_{lr}+p_{rl}+p_{ll}\right]/c\Omega[a,b]=2\cosh{\left[-q_1 al_1(b+l_2)\right]} e^{-1/2 q_1 l_1^2(b+l_2)}e^{-q_2 b l_2} \\
%+2\cosh{\left[-q_1 al_1(b-l_2)\right]}e^{-1/2 q_1 l_1^2(b-l_2)}e^{q_2 b l_2}       
%\end{eqnarray}

\section{Appendix B}

Before including induction effects, the dynamics for the gate variable is given by $\dot{b}=-b/\tau_{GT}$ where the time constant is related to the Onsager coefficient $L_{GT}$ and the generalized capacitance coefficient $g_{GT}^{-1}$ as $\tau_{GT}=1/L_{GT}g_{GT}$.
To determine the complete dynamics of the gate variable $b$, the following approach is used which has been successful in describing dissipative systems approaching equilibrium:
\begin{equation}
\langle \dot{b}(t')\rangle=\left\langle \dot{b}(t')e^{\Delta S(t'-t)/k_B}\right\rangle_0~,   \label{expS}          
\end{equation} 
where the brackets $\langle~~\rangle_0$ denote ensemble averaging over equilibrium states.  This approach has worked well for cases where kinetic coefficients are constant and when they vary linearly with gate variables, i.e., 1st order TI~\cite{Patitsas2014,Patitsas2015}.  However when the dependence is quadratic, the equilibrium averaging gives zero since $\langle b \rangle_0=0$.  Here a more refined type of averaging is required, the gate system still very close to equilibrium but with a non-zero average value.  Apart from this restriction all microstates will be sampled as equally as possible.  This is the essence of the exponential factor in Eq.~(\ref{expS}); configurations with greater multiplicities receive more weight in the statistical averaging, and the entire system is pushed towards equilibrium.
% mention total system entropy close to here
Thus I replace the brackets $\langle~~\rangle_0$ with the brackets $\langle~~\rangle_{ne}$, denoting near equilibrium.  This means that not all microstates will be accessed.  Instead all microstates are equally accessed, subject to the the restriction $\langle b \rangle\neq 0$.
For small time intervals, the change $\Delta S$ is small and 
\begin{equation}
\langle \dot{b}(t')\rangle=\frac{1}{k_B}\left\langle \dot{b}(t')\Delta S(t'-t)\right\rangle_{ne} \,.   \label{ne}          
\end{equation} 
%basically sample as many microstates as are possibly accessible given the restriction of $\langle b \rangle\neq 0$.
Integrating both sides of Eq.~(\ref{ne}) over the time interval $\Delta t_q$ gives the coarse-grained time derivative:
\begin{equation}
\bar{\dot{b}}=\frac{1}{k_B\Delta t}\int_{t}^{t+\Delta t}{dt'\langle\dot{b}(t')\Delta S(t'-t)\rangle_{ne}} \,. \label{coarse}
\end{equation} 
A bar is used to denote the coarse-graining and it is understood that the time step $\Delta t$ is very short but still much larger that the correlation time $\tau^*$ for the random force driving the fluctuations.  
The difference $\Delta S(t'-t)$ is replaced by the time integral of $\dot{S}_{T}$.  
The component of  $\dot{S}_{T}$ from to the gate alone results in the familiar term $L_{qq}X_{q}$ while $\dot{S}_{DR}$ is what supplies the more interesting induction term.
The contribution to $\bar{\dot{b}}$ that is exclusively thermodynamic induction is obtained by using only the $\alpha X_{DR}^2 b ^2 $ contribution to $\dot{S}_{DR}$.
Thus
\begin{equation}
\bar{\dot{b}}_{ind}=\frac{\alpha_q}{k_B\Delta t}\int_{t}^{t+\Delta t}{dt' \int_t^{t'} { dt''  \langle\dot{b}(t') X_{DR}^2 b^2(t'')\rangle_{ne}} } \,, \label{ind}
\end{equation} 
which can be re-expressed as
\begin{equation}
\bar{\dot{b}}_{ind}=\frac{\alpha X_{DR}^2}{ k_B\Delta t}\int_{t}^{t+\Delta t}{dt' \int_t^{t'} { dt'' \int_{-\infty}^{t''}{dt'''  \langle\dot{b}(t') \dot{b}(t''') b(t''')\rangle_{ne}}} } \,. \label{ind2}
\end{equation}
The correlation function $\langle\dot{b}(t') \dot{b}(t''') \rangle_{0}\equiv K_{qq}(t'-t''')$ is very small unless $|t'-t'''|< \tau^*$. 
Thus $\langle\dot{b}(t') \dot{b}(t''') b(t''')\rangle_{ne}$ is expected to be proportional to $K_{qq}(t'-t''')\langle b(t')\rangle_{ne}$.  When comparing to Eq.~(\ref{AppA} ) the constant of proportionality must be 2. 

%which gives Eq.~(0) without the result
%Added note May 26, 2018, this must be $\langle\dot{b}(t') \dot{b}(t''') b(t''')\rangle_{ne}\approx  2K_{qq}(t'-t''')\langle b(t')\rangle_{ne} $.
%This will probably be left out anyways.

Evaluating the time integrals gives $k_B L_{GT}\tau^*\Delta t$ so that (dropping the coarse-graining symbol)
\begin{equation}
\dot{b}_{ind}=2\alpha X_{DR}^2 L_{GT}\tau^* b~\, . \label{ind3}
\end{equation}
%similar analysis on the entropy difference for the gate gives the familiar Lqq Xq term.  Putting these together gives
Adding the induction in to the regular dynamics gives
\begin{equation}
\dot{b}=2\alpha X_{DR}^2 L_{GT}\tau^* b  -L_{GT}g_{GT}b~\, . \label{aqdot}
\end{equation}
%Equation~(\ref{aqdot}) constitutes the main result of this paper as it shows how unstable dynamics may occur in a purely dissipative system.
%
%end of Appendix B

%\bibliography{Bib17Oct2015}
\bibliography{Bib24May2019}

\end{document}